\documentclass[preprintnumbers,amsmath,amssymb,floatfix,10pt,prd,onecolumn,superscriptaddress,nofootinbib]{revtex4}
\usepackage{latexsym}
\usepackage{epsfig}
\usepackage{epstopdf}
\usepackage{graphicx}
\usepackage{amssymb}
\usepackage{amsmath}
\usepackage{amsfonts}
\usepackage{subfigure}
\usepackage{dcolumn}
\usepackage{bm}
\usepackage{color}
\usepackage{comment}
\usepackage[utf8]{inputenc}

\begin{document}
\title{\bf Optical Features of Rotating Black Hole with Nonlinear Electrodynamics}

\author{M. Zubair}
\email{drmzubair@cuilahore.edu.pk; mzubairkk@gmail.com}\affiliation{Department of Mathematics, COMSATS University Islamabad, Lahore Campus, Lahore, Pakistan}

\author{Muhammad Ali Raza}
\email{maliraza01234@gmail.com}\affiliation{Department of Mathematics, COMSATS University Islamabad, Lahore Campus, Lahore, Pakistan}

\author{Ghulam Abbas}
\email{ghulamabbas@iub.edu.pk}\affiliation{Department of Mathematics, The Islamia University of Bahawalpur, Pakistan}

\begin{abstract}
In this article, we considered the strong field approximation of nonlinear electrodynamics black hole and constructed its rotating counterpart by applying the modified Newman-Janis algorithm. The corresponding metric function in the strong field limit of the static black hole is identified in order to study the radius of photon sphere. However, the metric function for the rotating counterpart in the strong field limit is considered in order to study the horizon radius w.r.t spin parameter. We considered the Hamilton-Jacobi method to derive the geodesic equations for photon and constructed an orthonormal tetrad for deriving the equations for celestial coordinates in the observer's sky. Shadows, distortions and energy emission rates are investigated and the results are compared for different values of nonlinear electrodynamics parameter, charge and spin. It is found that the presence of the nonlinear electrodynamics parameter affects the shape and size of the shadows and thus the distortion in the case of rotation. It is also found that the nonlinearity of electrodynamics diminishes the flatness in the shadow due to the effect of spin and other parameters.
\end{abstract}
\maketitle
\date{\today}

\section{Introduction}
General Relativity (GR) predicts regions of ultra strong gravity known as black holes (BHs) having immense curvature. The photons will deviate from its straight path while moving in the gravitational field of the BHs and other massive objects. Such concept was verified by Arthur Eddington and his teammates in $1919$ \cite{1} that ultimately verified GR. It is called gravitational lensing which is a general term assigned to the study of effects that are caused by the deflection of photons and an outstanding progress has been made in the field over the years \cite{2,3,4,5,6,7,8,9}.

The photons moving close enough to the BH will get trapped, otherwise they will scatter away. This trapping region is called photon sphere in which photons orbit the BH in circular paths. The scattered photons present in the unstable circular orbits will reach our eyes giving a glowing image but falling in photons will be lost, giving us a dark $2$D image called shadow. The falling in trajectory of the photons in the outermost photon region and the scattering trajectory in the innermost region will define the boundary of the shadow \cite{10}. Usually the shape of the static BH shadows is different from the rotating ones due to the fact that the photons may move in any direction around a BH in the photon region of certain width. Without loosing generality, photons in the both extreme orbits can be considered moving in opposite directions. Thus, one side of the shadow appears flattened as compared to the other. Shadow of Kerr BH is one such example \cite{10,11}.

The shadow study remains one of the foremost topics related to the BHs because whenever a BH solution is discovered the first arising question is that what would likely be its physical appearance? Some earlier studies related to visual appearance of BHs are \cite{12,13,14,15,16,17}. In these studies, the visual appearance of the BH was termed with various names such as optical appearance, escape cone, cross section, cone of gravitational capture of radiation etc. Over the years, some mathematical techniques \cite{18,19,20} were developed to study the shadows analytically. The authors of \cite{18} and \cite{19} investigated the shadows by considering the observer at a position $(r_{0},\vartheta_{0})$. Using the orthonormal tetrad at this position, the authors have constructed the mathematical framework for celestial coordinates in the observer's sky. This method is widely applicable for both distant and nearer observers. Bardeen \cite{20} studied the geodesics for Kerr BH by considering two impact parameters. He developed his own method for studying the optical properties in the vicinity of a BH. However, this method is applicable only for a distant observer. Despite its limitation, Bardeen's method is the most commonly used procedure for the shadow study, see \cite{21,22,23,24,25,26,27,28,29,30}. Atamurotov et al. \cite{25} investigated the shadows of Kerr-Newman BH immersed in perfect fluid dark matter by considering the Bardeen's method. They found that the shadow is flattened by increasing the value of spin and the size of the shadow decreases with increase in the values of charge and perfect fluid dark matter.

By the discovery of more general BH solutions, it became difficult to deal with the mathematical structure manually. As a result, various software tools were used in order to suggest the possible appearance of such complicated BH solutions. Some notable studies are \cite{31,32,33,34,35,36,37,38,39,40,41,42,43}. Hioki and Maeda \cite{32} calculated the shadows and the related observables by the use of software, especially the contour plots therein. The BH solutions formulated in various theories of gravity depend upon different BH parameters. These parameters have different sensitivity level. The shadows are highly influenced by more sensitive BH parameters. As said above, spin parameter flattens the shadow on one side for an equatorial observer and the presence of cosmological constant, perfect fluid dark matter, quintessence etc. have a significant impact on the size and shape of the shadows, for more details see \cite{44,45,46,47,48,49}. Haroon et al. \cite{50} studied the effects of perfect fluid dark matter $(\alpha)$ and cosmological constant $(\Lambda)$ on the shadows of rotating BH. They found that for negatively increasing $\alpha$, the shadow size increases while the shadow shrinks for positively increasing $\alpha$. Moreover, the distortion due to high spin value diminishes for $\alpha \gtrsim 0.8$. The shadow size increases for anti-de Sitter case and decreases for de Sitter case.

In Maxwell's electrodynamics, there exist a singularity at the position of the point charge and has an infinite self-energy. To overcome this problem, Born and Infeld \cite{51} developed a nonlinear electromagnetic field. Motivated by this, the coupling of GR and Born-Infeld field has been studied in order to deal with the singularity problem of the BHs along with some other properties, see \cite{52,53,54,55}. Demianski \cite{52} derived an asymptotically flat solution which becomes regular spacetime when the internal mass is considered zero. Cai et al. \cite{54} studied the BH solution and its thermodynamic properties in Born-Infeld theory. The coupling of GR and some other nonlinear electrodynamics models have also been studied providing us with some useful results in BH physics \cite{56,57,58,59,60}. Javed et al. \cite{60} considered a magnetized nonlinear electrodynamics BH and studied the effect of nonlinear electrodynamics parameter $\beta$ on the deflection angle in the vicinity of BH. Kruglov \cite{61} constructed a BH solution with nonlinear electromagnetic field. This BH solution is asymptotically Reissner-Nordstr\"{o}m and the electric field has finite value at the origin which does not possess singularity at $r=0$. Recently, Uniyal et al. \cite{62} considered the same solution developed by Kruglov and studied the photon sphere and shadow in the weak and strong field limits.

In this paper, we consider the nonlinear electrodynamics BH solution as in Refs. \cite{61,62} and apply the modified Newman-Janis algorithm \cite{63,64} to the effective metric in the strong field approximation of the BH. We work for the horizons, shadows and related physical observables to examine the effect of nonlinearity of electromagnetic field. The paper is presented as: In section $2$, the nonlinear electrodynamics BH solution is presented. The approximated metric in the strong field limit is considered for further analysis and the radius of photon sphere is studied for the non-rotating case. In section $3$, rotating metric is constructed and the corresponding horizon radii are studied. In section $4$, the governing equations for shadows are developed using the Hamilton-Jacobi formalism and the method of orthonormal tetrads. The shadows, distortions and energy emission rates are presented for the observer at different locations. We summarize the results in the last section. Note that the units $G=c=1$ have been used.

\section{$4$D Nonlinear Electrodynamics Black Hole}
We start by the review of the non-rotating nonlinear electrodynamics BH. The gravitational action containing the Lagrangian of nonlinear electromagnetic field \cite{61,62} is given as
\begin{equation}
S_G=\int d^{4}x\sqrt{-g}\bigg[\frac{R}{2\kappa^{2}}+\mathcal{L}_{em}\bigg], \label{1}
\end{equation}
where $g=det(g_{\mu\nu})$, $R$ is Ricci scalar, $\kappa^{-1}$ is the reduced Planck mass and $\mathcal{L}_{em}$ is the nonlinear electromagnetic Lagrangian \cite{65} given as
\begin{equation}
\mathcal{L}_{em}=-\frac{\mathcal{F}}{2\beta\mathcal{F}+1}, \label{2}
\end{equation}
such that $\beta$ is the corresponding parameter of nonlinear electromagnetic field with the dimension of $(length)^4$ and is related to the upper bound of the electric field, $\mathcal{F}=\frac{F_{\mu\nu}F^{\mu\nu}}{4}$ and $F_{\mu\nu}$ is the Maxwell tensor. The Lagrangian (\ref{2}) corresponds to classical linear electrodynamics when $\beta=0$. The nonlinearity of electrodynamics in various forms, coupled with gravity can be useful in achieving some of the unanswered questions. Furthermore, for the possible quantum gravity corrections to Maxwell's electrodynamics, the parameter $\beta$ is introduced as in \cite{65}. The function (\ref{2}) is formulated in such a way that the correspondence principal is not broken and the model works efficiently with usual dielectric permittivity $\varepsilon=1$ and magnetic permeability $\mu=1$. Hence, the action (\ref{1}) in \cite{61,62} is the possible coupling of the function (\ref{2}) with GR. In this model, the energy momentum tensor has non-zero trace and at the origin, the electric field has a finite value without singularities, for more details, see \cite{61,65}. For a $4$D spacetime in spherical symmetry, we have
\begin{equation}
ds^{2}=-f(r)dt^{2}+\frac{dr^{2}}{f(r)}+r^{2}(d\vartheta^{2}+\sin^{2}\vartheta d\varphi^{2}), \label{3}
\end{equation}
the solution turns out to be \cite{61,62}
\begin{equation}
f(r)=1-\frac{2M}{r}+\frac{Q^{2}}{r^{2}}-\frac{C^{2}\kappa^{2}}{2r^{2}}+\frac{C^{2}\kappa^{2}}{30r^{2}}(5\zeta^{3}-22\zeta^{2}+32\zeta), \label{4}
\end{equation}
where $M$, $Q$ and $C$ are mass, charge and integration constant, respectively. Here,
\begin{equation}
\zeta=\frac{12\sqrt{3}\sqrt{\beta}r^{2}-\beta C\lambda^{3/4}}{12\beta C\lambda^{1/4}}, \label{5}
\end{equation}
whereas
\begin{eqnarray}
\lambda&=&\frac{6\sqrt[3]{6}r^{2}(\sqrt[3]{2}\beta^{2/3}\sqrt[3]{C}\gamma^{2/3}-8\sqrt[3]{3}\beta C)}{\beta^{4/3}C^{5/3}\sqrt[3]{\gamma}}, \label{6} \\
\gamma&=&\sqrt{3}\sqrt{256\beta C^{2}+27r^{4}}+9r^{2}. \label{7}
\end{eqnarray}

Keeping in view the complexity of $\zeta$, the simple forms of metric function $f(r)$ in strong and weak field limits are obtained by applying series expansion for $r\rightarrow0$ and $r\rightarrow\infty$ respectively, such that
\begin{eqnarray}
f(r)_s&=&1-\frac{2M}{r}+\frac{Q^{2}}{r^{2}}-\frac{C^{2}\kappa^{2}}{2r^{2}}+\frac{16C^{3/2}\kappa^{2}}{15\beta^{1/4}r}, \label{8} \\
f(r)_w&=&1-\frac{2M}{r}+\frac{Q^{2}}{r^{2}}-\frac{\beta C^{4}\kappa^{2}}{10r^{6}}. \label{9}
\end{eqnarray}

Since, we aim to study the horizon, photon sphere and shadow, so it is impossible to proceed with the metric function (\ref{4}) due to its complication and the required mathematical structure. Also, the horizons, photon spheres and shadows are strong field phenomena that exist only in the vicinity of the BHs. Thus, to avoid the complexity of the metric function (\ref{4}), we will consider the metric function (\ref{8}) explicitly in further analysis, treating the metric as an effective metric. This effective metric, when stretched out again in the entire universe, becomes asymptotically flat as $r\rightarrow\infty$. It will raise some extra sources and will not satisfy the field equations exactly the same way as the metric (\ref{4}) because both of these metrics are not exactly equal. Due to this reason, their properties will also be slightly different. However, without caring for the difference in properties, we will proceed with the effective metric in order to explore the effect of nonlinear electrodynamics in the strong field limit which is the intent of this study.

As we move closer to the horizon, the effect of nonlinear electrodynamics is highly impactful and can not be removed. Apparently, it seems that the effect of nonlinear electrodynamics can be removed in the strong field region by considering the limit $\beta\rightarrow\infty$ in $f(r)_s$. However, this is physically not a realistic limit, since under this limit, the Lagrangian (\ref{2}) will be vanished. Hence, under a finite value of the parameter $\beta$, if we consider $C=0$, we will get the Reissner-Nordstr\"{o}m metric which is again not a useful limit. Hence, we deduce that nonlinear electrodynamics effect can not be removed in the vicinity of the BH.

Note that the index $s$ occurring as the superscript or subscript throughout the discussion corresponds to the functions and variables in the strong field limit. As mentioned above, the symbol $C$ is just a dimensionless constant and has no physical meaning discussed in the Ref. \cite{61,62}. So, we will fix its value in our work because we mainly aim at the parameter $\beta$. However, the constant $C$ plays an important role in distinguishing the metric in strong field limit from Reissner-Nordstr\"{o}m metric because inserting $C=0$ in Eq. (\ref{8}), we will get the standard Reissner-Nordstr\"{o}m metric. So, for a rigorous analysis, the non-zero value of $C$ must be considered. To avoid the imaginary numbers, the negative values of $C$ can not be considered. Hence, we will consider $C=1$ in our calculations in the same way as we consider $M=1$ quite often for simplicity.

In the function $f(r)_s$, we can see that there exist two extra terms of $\mathcal{O}(\frac{1}{r})$ and $\mathcal{O}(\frac{1}{r^2})$ and apparently the metric (\ref{8}) seems identical to the Reissner-Nordstr\"{o}m metric by shifting the constants and parameters in $f(r)_s$. However, the terms containing $\beta$, $C$ and $\kappa$ make this metric distinct from the Reissner-Nordstr\"{o}m metric because of their physical nature lying therein. Since, the Lagrangian (\ref{2}) consists a function of Maxwell's tensor but with the presence of $\beta$, defining the nonlinear electromagnetic theory. Then the resulting function $f(r)_s$ depends upon $M$, $Q$, $\beta$, $C$ and $\kappa$. These parameters and constants are responsible for defining the strong field metric in nonlinear electromagnetic theory. However, the Maxwell's Lagrangian results in Reissner-Nordstr\"{o}m BH whose metric function depends upon $M$ and $Q$ only and is linearly electrically charged. Whereas, the metric under consideration consists of charge with strong nonlinear effects. As we know that the trace of energy-momentum tensor is zero for the Maxwell's Lagrangian that gives Reissner-Nordstr\"{o}m BH. However, for the nonlinear electrodynamics case, the trace of energy-momentum tensor is non-zero for the Lagrangian (\ref{2}). This non-zero value of the trace contains $\beta$ and further contains $C$ in the function $h'(r)$ as given in Refs. \cite{61,62}. By considering $\beta=0$, the trace becomes zero. Also, by taking the value $C=0$ in the trace and in the metrics (\ref{4}) and (\ref{8}), the trace becomes zero and the metrics reduce to Reissner-Nordstr\"{o}m metric. Hence, the non-zero finite values of $\beta$ and $C$ define our metric that is distinguishable from the Reissner-Nordstr\"{o}m metric i.e. by the comparison of the trace, the strong field metric is distinguishable from the Reissner-Nordstr\"{o}m metric.

Suppose that after the shifting, we get the parameters $m=M-\frac{8C^{3/2}\kappa^2}{15\beta^{1/4}}$ as effective mass and $q=\sqrt{Q^2-\frac{C^2\kappa^2}{2}}$ as effective charge in $f(r)_s$. Then the resulting metric would appear as Reissner-Nordstr\"{o}m metric with parameters $m$ and $q$ as said above. This would be significant only if we use the parametric values of $m$ and $q$ with a physical description. However, there is no physical description of $m$ and $q$ apart from calling them as effective mass and effective charge, respectively. Moreover, we can not use the values of $m$ and $q$ independently in the calculations because if we put $m=0.5$ and $q=0.6$, these are actually the combination of parametric values of $M$, $Q$, $\beta$, $C$ and $\kappa$. So, if we have to use the values of $M$, $Q$, $\beta$, $C$ and $\kappa$ at the end, then considering $m$ and $q$ is of no use. Else, if we use the values of $m$ and $q$, then the required explanation of the results would be related to $M$, $Q$, $\beta$, $C$ and $\kappa$ which is inconsistent with what is used in the calculations. Furthermore, using $m$ and $q$ would suppress the actual objective of this study because we aim to focus on the effect of nonlinear electrodynamics parameter on the photon sphere, horizon and shadow. As said earlier, the nonlinear electrodynamics effects can not be removed or neglected in the strong field region and the nonlinear electrodynamics is not the characteristics of Reissner-Nordstr\"{o}m metric. Hence, the metric (\ref{8}) is different from the Reissner-Nordstr\"{o}m metric for all of the above mentioned reasons. Later, this fact will also be demonstrated through the results of photon sphere, horizon and shadow.

To discuss the radius of photon sphere in the strong field limit, we assume an observer near the BH. Then the general condition for the radius of photon sphere is \cite{10}
\begin{equation}
\frac{d}{dr}\bigg(\frac{r^{2}}{f(r)_s}\bigg)\bigg|_{r=r^s_{p}}=0, \label{10}
\end{equation}
which gives
\begin{equation}
r^s_{p}=\frac{3}{2}\bigg(M-\frac{8C^{3/2}\kappa^{2}}{15\beta^{1/4}}\bigg)+\frac{1}{2}\sqrt{9\bigg(M-\frac{8C^{3/2}\kappa^{2}}{15\beta^{1/4}}\bigg)^2-8Q^2+4C^{2}\kappa^{2}}. \label{11}
\end{equation}

Figure $\textbf{1}$ presents the plots for the radius of photon sphere vs $\beta$ and $Q$ and shows a comparison of results with Reissner-Nordstr\"{o}m BH. Note that, for all plots, we have used $M=1$, $C=1$ and $\kappa=1$ for simplicity. We have kept the same values and variations for $\beta$ for a rigorous comparison. The negative values of $\beta$ have not been considered because there exist no photon sphere for $\beta<0$ in the strong field limit. So, $\beta$ is kept positive for all calculations. Also, $\beta$ being less sensitive as compared to other parameters, is not varied with small step size in the intervals such as $(0,1)$. Instead, we vary it in the interval $(0,100]$ keeping in view the sensitivity level. Moreover, it is still possible to do calculations for $\beta\in(0,1)$. For example, there will be finite values of radius of photon sphere for $\beta\in(0,1)$. However, the results will not be convergent to $r_p^s=3$ (Schwarzschild BH) for small variation in the values of $\beta$ and for $Q=0$. For a comprehensive analysis, it is important that the values of radius of photon sphere should approach the value of Schwarzschild BH. So, as $\beta\gg100$ and $Q=0$, the value of radius of photon sphere approaches $3$. Whereas, for a small step size of $\beta$, the radius of photon sphere will be approximately around $1.5$, which is not a significant value to compare the results.

In the left plot, it can be seen that when $\beta\rightarrow0$, the radius $r^s_p$ drops asymptotically for different values of charge $Q$ for nonlinear electrodynamics case. However, for $Q=1$, minimum value of $r^s_{p}\sim1$. Also, the radius increases rapidly till $\beta\sim10$ and for other values of $\beta$, radius increases slowly and approaches the corresponding values of Reissner-Nordstr\"{o}m BH shown with dashed curves. Moreover, the outermost curve corresponds to $Q=0$ and by increasing $Q$, the curves are shifted inwards. The dashed curves show a constant value of radii for each fixed value of $Q$ because the Reissner-Nordstr\"{o}m BH has no dependence upon $\beta$. In the right plot, it is found that the radius decreases as charge increases for each curve. Also the curves are shifted outwards for increasing $\beta$. As we know that certain parameters weaken the gravity and hence the photon regions become smaller by increasing the values of those parameters. Here, such behavior is also shown by the charge parameter $Q$ as if it is responsible for weakening the gravity. The outermost orange curve corresponds to the Reissner-Nordstr\"{o}m BH and all of the other curves approach it under the limit $\beta\rightarrow\infty$. However, under this limit, these curves will not exactly match the curve for Reissner-Nordstr\"{o}m BH unless $C=0$.

It is obvious that our metric being different from the standard Reissner-Nordstr\"{o}m metric generated the distinct results for the radii of photon sphere. On observing the plots, we see that the plots are dependent upon $\beta$ which is not a characteristics of a standard Reissner-Nordstr\"{o}m metric. The curves in the left plot for our metric are no way near to the curves for Reissner-Nordstr\"{o}m metric because the solid curves have positive slopes while the dashed curves are constant due to no dependence upon $\beta$. In the right plot, for the Reissner-Nordstr\"{o}m metric, the radius is observed equal to $3$ for $Q=0$ and then gradually drops as $Q$ increases. The other curves are distinct but approach the curve for Reissner-Nordstr\"{o}m metric when $\beta$ is large enough. This makes our metric distinguishable from the Reissner-Nordstr\"{o}m metric because otherwise there would not be such a large difference in radii of photon sphere. In other words, the Reissner-Nordstr\"{o}m metric is a special limiting case of the metric (\ref{8}).

\section{Rotating Metrics for $4$D Nonlinear Electrodynamics Black Hole}

For the derivation of rotating counterparts, Newman-Janis algorithm \cite{66,67} is usually applied to the static BHs in GR. However, it has also been applied to some static BH solutions in non-GR modified gravity theories \cite{68,69,70}. We know that by applying the Newman-Janis algorithm to any non-GR BH solution raises pathologies in the resulting rotating solution \cite{71}. It means that the source in GR must be the same for both a static BH and its rotating counterpart, i.e., vacuum (charge) for both Schwarzschild (Reissner-Nordstr\"{o}m) and Kerr (Kerr-Newman) BHs. However, in the modified gravity theories, there exist some extra sources for the rotating BHs derived by Newman-Janis algorithm. To encounter this problem, Azreg-A\"{\i}nou \cite{63,64} derived the modified form of Newman-Janis algorithm which is a non-complexification method and also assumes arbitrary metric functions, whose values are to be determined. This makes the method easy to deal with and also it is applicable to more variety of BHs in modified theories as well as GR resulting in the exact rotating BH solutions. It is also successfully applied to derive the imperfect fluid rotating counterparts and generic rotating regular BH solutions in Boyer-Lindquist coordinates $(t,r,\theta,\phi)$ \cite{29,72,73}. Hence, the modified Newman-Janis algorithm being a general method can also be applied to the BH solutions within GR. With this motivation and as we are working in GR, we applied this method to the metric (\ref{8}). Since, we consider this metric explicitly as an effective metric, so we will derive its rotating counterpart by the direct application of the modified NJA. We start by considering Eddington-Finkelstein coordinates $(u,r,\vartheta,\varphi)$ such that
\begin{eqnarray}
dt=du+\frac{dr}{f(r)_s}. \label{12}
\end{eqnarray}
By using the transformation (\ref{12}), we get
\begin{equation}
ds^{2}=-f(r)_sdu^{2}-2dudr+r^{2}(d\vartheta^{2}+\sin^{2}\vartheta d\varphi^{2}). \label{13}
\end{equation}
The null tetrad for the metric can be written as
\begin{eqnarray}
l^a&=&\delta^a_r, \label{14}\\
n^a&=&\delta^a_u-\frac{f(r)_s}{2}\delta^a_r, \label{15}\\
m^a&=&\frac{1}{\sqrt{2}r}\bigg(\delta^a_\vartheta + \frac{i}{\sin{\vartheta}}\delta^a_\varphi\bigg).  \label{16}
\end{eqnarray}
Since, the above null tetrad gives the metric tensor as
\begin{figure}[t]
\begin{center}
\subfigure{
\includegraphics[height=5.5cm,width=8cm]{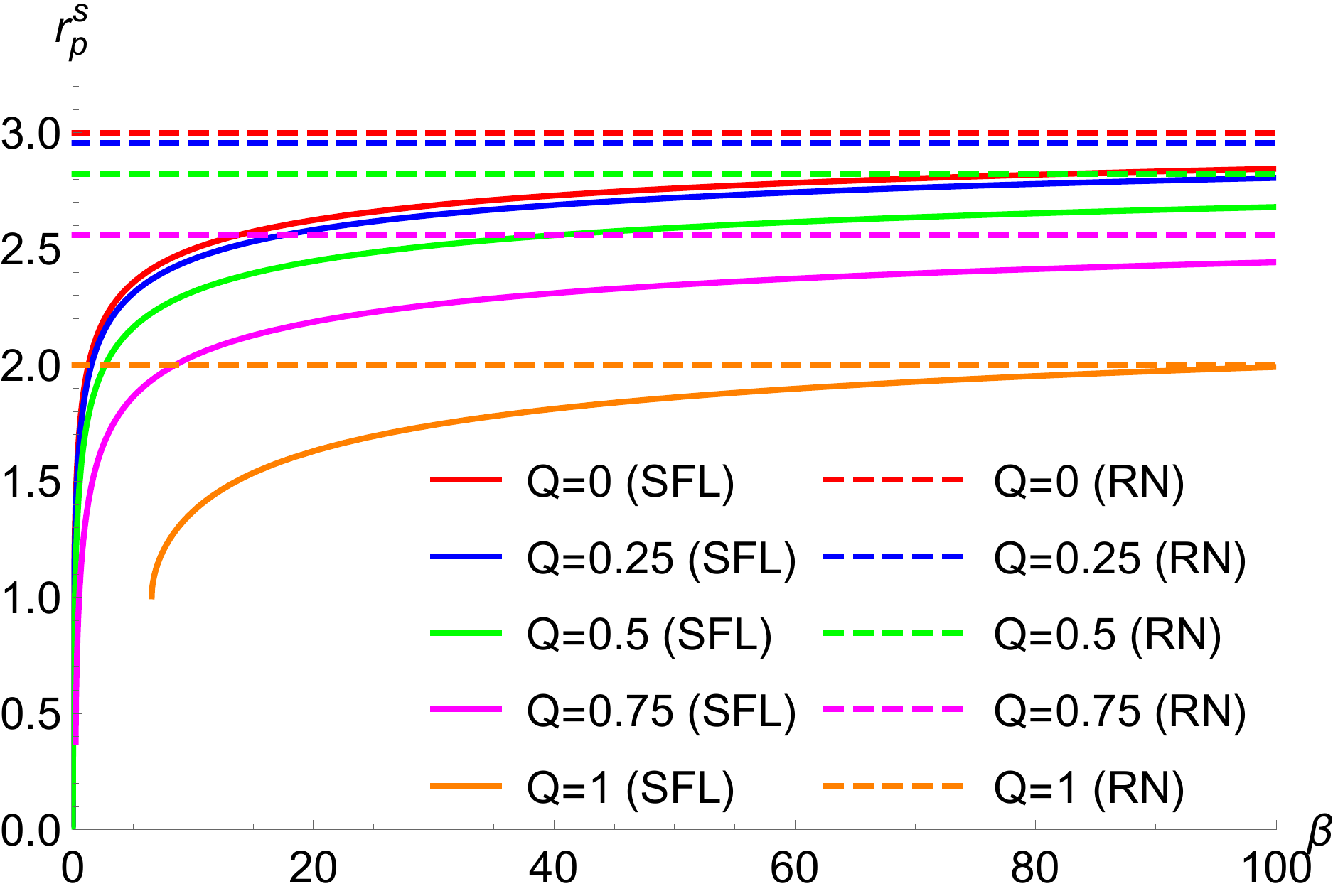}}
~~~~~~~~~
\subfigure{
\includegraphics[height=5.5cm,width=8cm]{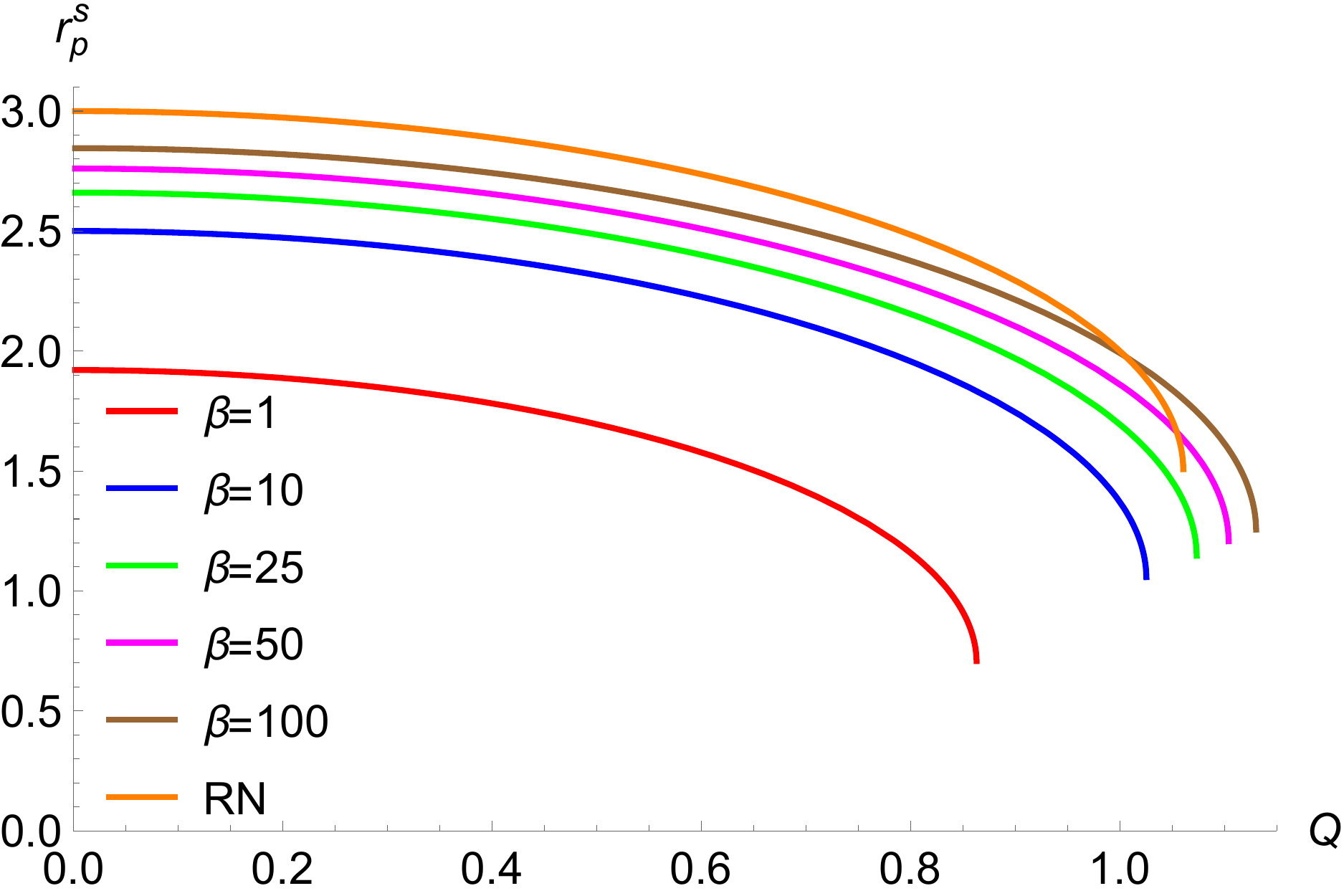}}
\end{center}
\caption{Variation of radius of photon sphere in strong field w.r.t $\beta$ and $Q$ for the static case. RN and SFL stand for Reissner-Nordstr\"{o}m metric and the metric in strong field limit, respectively.}
\end{figure}
\begin{equation}
g^{ab}=-l^an^b-n^al^b+m^a\bar{m}^b+\bar{m}^am^b, \label{17}
\end{equation}
where $\bar{m}^a$ is the complex conjugate of $m^a$. Furthermore, the null tetrad $(l^a,n^a,m^a,\bar{m}^a)$ obeys the following conditions:
\begin{eqnarray}
l_al^a=n_an^a=m_am^a=\bar{m}_a\bar{m}^a&=&0, \label{18}\\
l_am^a=l_a\bar{m}^a=n_am^a=n_a\bar{m}^a&=&0, \label{19}\\
-l_an^a=-l^an_a=m_a\bar{m}^a=m^a\bar{m}_a&=&1. \label{20}
\end{eqnarray}
The complex coordinate transformations for the rotation in $(u,r)$-plane with the spin parameter $a$ are given by
\begin{eqnarray}
u'&\rightarrow& u-ia\cos{\vartheta}, \label{21}\\
r'&\rightarrow& r+ia\cos{\vartheta}. \label{22}
\end{eqnarray}
The new undetermined metric functions are given as
\begin{eqnarray}
f(r)_s&\rightarrow& \mathcal{A}, \label{23}\\
r^2&\rightarrow& \mathcal{B}, \label{24}
\end{eqnarray}
where $\mathcal{A}$ and $\mathcal{B}$ are the functions of $r$, $a$ and $\vartheta$. Using the complex transformations and the new metric functions, we get the transformed tetrad with spin as
\begin{eqnarray}
l'^{a}&=&\delta^a_r, \label{25}\\
n'^{a}&=&\delta^a_u - \frac{\mathcal{A}(r,a,\vartheta)}{2}\delta^a_r, \label{26}\\
m'^{a}&=&\frac{1}{\sqrt{2\mathcal{B}}}\bigg((\delta^a_u-\delta^a_r)ia\sin{\vartheta}+\delta^a_\vartheta + \frac{i}{\sin{\vartheta}}\delta^a_\varphi\bigg),  \label{27}
\end{eqnarray}
where prime denotes the transformed tetrad. Now, using this tetrad and the definition of metric tensor in terms of tetrad, the metric in Eddington-Finkelstein coordinates becomes
\begin{eqnarray}
ds^{2}&=&-\mathcal{A}du^{2}-2dudr+2a(\mathcal{A}-1)\sin^{2}\vartheta dud\varphi+2a\sin^{2}\vartheta drd\varphi+\mathcal{B}d\vartheta^{2} \nonumber \\
&&+\sin^{2}\vartheta(\mathcal{B}-(\mathcal{A}-2)a^2\sin^{2}\vartheta)d\varphi^{2}. \label{28}
\end{eqnarray}
We have dropped the primes in above equation for simplicity. In the last step, we transform the above metric into the Boyer-Lindquist coordinates by choosing the transformations as
\begin{eqnarray}
du&=&dt+\Gamma(r)dr, \label{29}\\
d\varphi&=&d\varphi'+\Xi(r)dr, \label{30}
\end{eqnarray}
with
\begin{eqnarray}
\Gamma(r)&=&-\frac{a^2+r^2}{a^2+r^2f(r)_s}, \label{31}\\
\Xi(r)&=&-\frac{a}{a^2+r^2f(r)_s}. \label{32}
\end{eqnarray}
Then by choosing
\begin{eqnarray}
\mathcal{A}=\frac{a^2\cos^2{\vartheta}+r^2f(r)_s}{\mathcal{B}}, ~~~ \mathcal{B}=r^2+a^2\cos^2{\vartheta}, \label{33}
\end{eqnarray}
we get
\begin{eqnarray}
ds^{2}&=&-\bigg(\frac{\Delta_s-a^2\sin^2{\vartheta}}{\rho^2}\bigg)dt^{2}+\frac{\rho^2}{\Delta_s}dr^2+\rho^2d\vartheta^{2}+\frac{\sin^{2}\vartheta}{\rho^2}\bigg(\bigg(r^2+a^2\bigg)^2-\Delta_s a^2\sin^{2}\vartheta\bigg)d\varphi^{2} \nonumber \\
&&+\frac{2a\sin^{2}\vartheta}{\rho^2}\bigg(\Delta_s-a^2-r^2\bigg)dtd\varphi, \label{34}
\end{eqnarray}
where
\begin{eqnarray}
\Delta_s&=&a^2+r^2f(r)_s=r^2-2Mr+a^2+Q^2-\frac{C^{2}\kappa^{2}}{2}+\frac{16C^{3/2}\kappa^{2}}{15\beta^{1/4}}r, \label{35}\\
\rho^2&=&r^2+a^2\cos^2{\vartheta}. \label{36}
\end{eqnarray}

The metric (\ref{34}) is the rotating counterpart for the metric (\ref{8}). In general, the rotating solution can be verified as an exact solution by satisfying the field equations in the same way as satisfied by the non-rotating solution. However, in our case, the metric (\ref{8}) is an approximation of the metric (\ref{4}) so it will not satisfy the field equations as satisfied by the metric (\ref{4}). Instead, the metric (\ref{8}) will satisfy a particular set of field equations that are unknown. Thus, the rotating metric (\ref{34}) will also not satisfy the field equations corresponding to the metric (\ref{4}). But, according to the properties of the modified Newman-Janis algorithm, the metric (\ref{34}) will satisfy the unknown set of field equations that are satisfied by the static metric (\ref{8}). Thus, the metric (\ref{34}) is the exact rotating counterpart of the metric (\ref{8}) and can be treated as the approximate rotating counterpart of the metric (\ref{4}). Moreover, if we consider $a=0$ in the metric (\ref{34}), we obtain the static metric (\ref{8}). It has been followed by Kumar and Ghosh \cite{29} that the rotating solution derived by modified Newman-Janis algorithm can be regarded as the most appropriate rotating solution despite the fact that it may not satisfy the field equations as satisfied by its non-rotating counterpart. Hence, in the same way, the metric (\ref{34}) can be regarded as the most appropriate rotating counterpart for the approximated metric (\ref{8}). This rotating metric can also be treated as an effective metric.

In order to study the horizon structure in strong field limit, we solve the equation $\Delta_s=0$, which gives
\begin{equation}
r^s_h=1-\frac{8}{15\beta^{1/4}}\pm\sqrt{\bigg(1-\frac{8}{15\beta^{1/4}}\bigg)^2-a^2-Q^2+\frac{1}{2}}. \label{37}
\end{equation}

The real values of horizon for different parametric values of $a$, $Q$ and $\beta$ depend upon the non-negativity of the discriminant in above equation. We aim for the horizon radius w.r.t $a$, so the discriminant is non-negative for particular values of $Q$ and $\beta$ for each value of $a$. Suppose that for some values of $Q$ and $\beta$, $\sqrt{\alpha-a^2}$ must be real such that $\alpha-a^2\geq0$. This implies that $a^2\leq\alpha$ and hence $a~\epsilon~[-\sqrt{\alpha},\sqrt{\alpha}]$. As we are dealing with only non-negative values of $a$, so $a~\epsilon~[0,\sqrt{\alpha}]$. Hence, we get
\begin{equation}
a~\epsilon~\bigg[0,\sqrt{\bigg(1-\frac{8}{15\beta^{1/4}}\bigg)^2-Q^2+\frac{1}{2}}\bigg]. \label{38}
\end{equation}

The horizon radius is plotted w.r.t $a$ in Fig. $\textbf{2}$ for different values of $Q$ and $\beta$. For the comparison, the results for the Reissner-Nordstr\"{o}m metric are also shown. The left panel shows that for $Q=0$ and $Q=0.5$, there does not exist Cauchy horizon for $a\lesssim 0.7$ and $a\lesssim 0.5$ respectively, only event horizon exists in this region. For $a\gtrsim 0.7$ and $a\gtrsim 0.5$, Cauchy horizon also exists and for $a\approx0.7$ and $a\approx0.5$ central singularity exits. However, for $Q=1/\sqrt{2}$, both Cauchy and event horizons are formed for all allowable values of $a\neq0$ and for $a=0$ there exists central singularity for each observed curve. Moreover, the curves are shifted inwards with decreasing $\beta$, which is precisely due to the effect of nonlinear electrodynamics and other parameters. The event horizon decreases whereas Cauchy horizon increases with increasing $a$ for each curve and each value of $Q$. The orange curves show the horizon radii for the Reissner-Nordstr\"{o}m metric. In the right panel, we can see that for $Q=0.8$, the Cauchy horizon is formed for all values of $a$ and has no central singularity. Whereas, for other curves corresponding to different values of $Q$, there does not exist Cauchy horizon for some values of $a$, only event horizon exists in this region. Then we can identify the values of $a$ where central singularity exists and for all greater values of $a$, the Cauchy horizon exists. Moreover, for each $\beta$, the curves are shifted inwards for increasing $Q$. It is shown that the event horizon decreases whereas Cauchy horizon increases with increasing $a$ for each curve corresponding to $Q$ and each value of $\beta$. The results for the Reissner-Nordstr\"{o}m metric are plotted with the dashed curves.

The metric (\ref{34}) and the Kerr-Newman metric depict very different results for the horizon radii. This is due to the fact that the Kerr-Newman metric depicts two horizons for all parametric values. However, in our case, we observe single horizon for some cases where the effect of $\beta$ has not been vanished. For $Q\geq1/\sqrt{2}$, the two horizons are observed for given values of $\beta$ and $a$. This is because the effect of $\beta$ and the extra terms of nonlinear electrodynamics vanish in the equations. The values of horizon radii are also very different for each particular set of $Q$ and $\beta$ w.r.t $a$. In this way, due to the different results, the metric (\ref{34}) is distinguishable from the Kerr-Newman metric and hence the metric (\ref{8}) is distinguishable from the Reissner-Nordstr\"{o}m metric.
\begin{figure}[t]
\begin{center}
\subfigure{
\includegraphics[height=7.5cm,width=8cm]{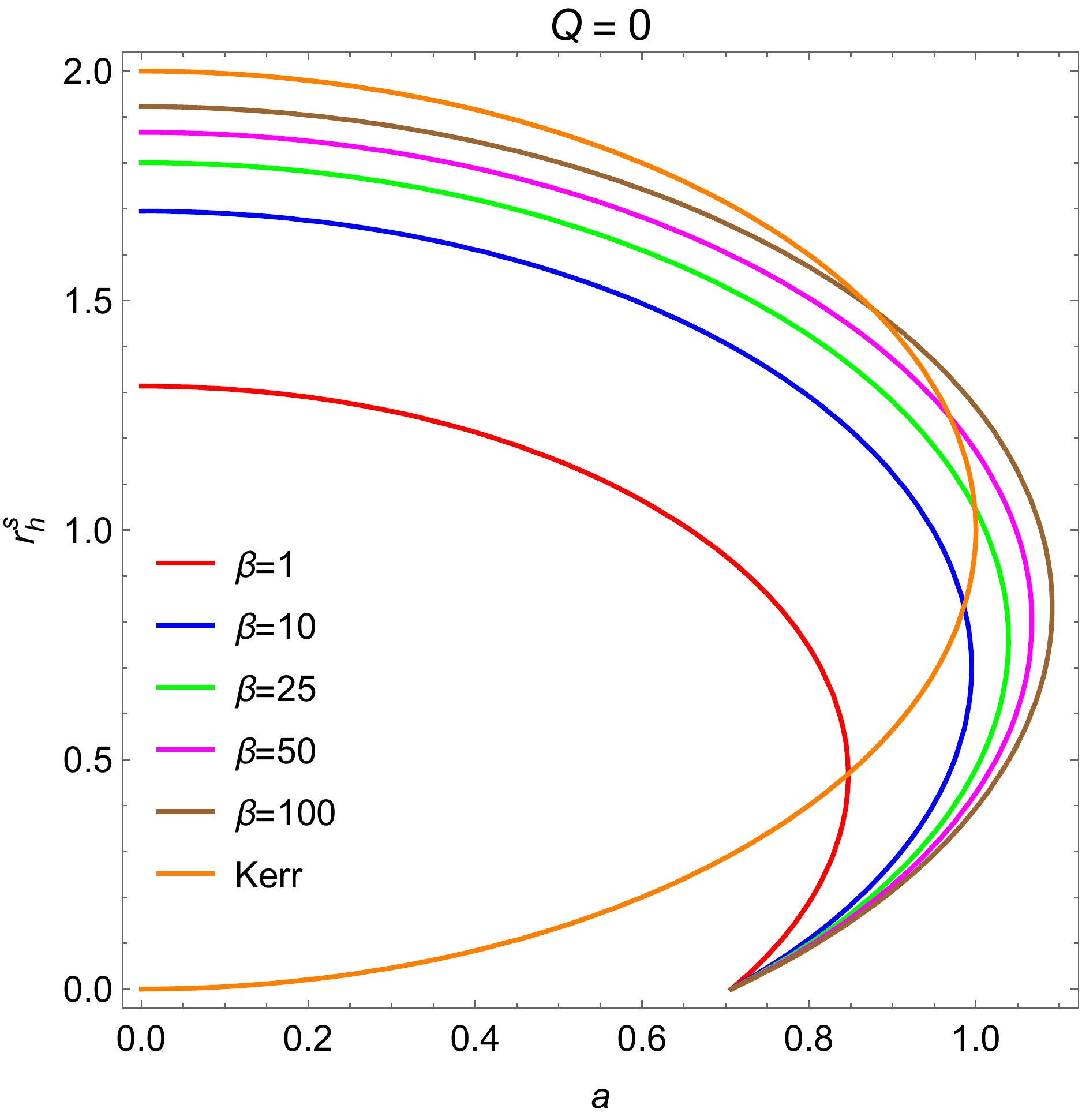}}
~~~~~~~
\subfigure{
\includegraphics[height=7.5cm,width=8cm]{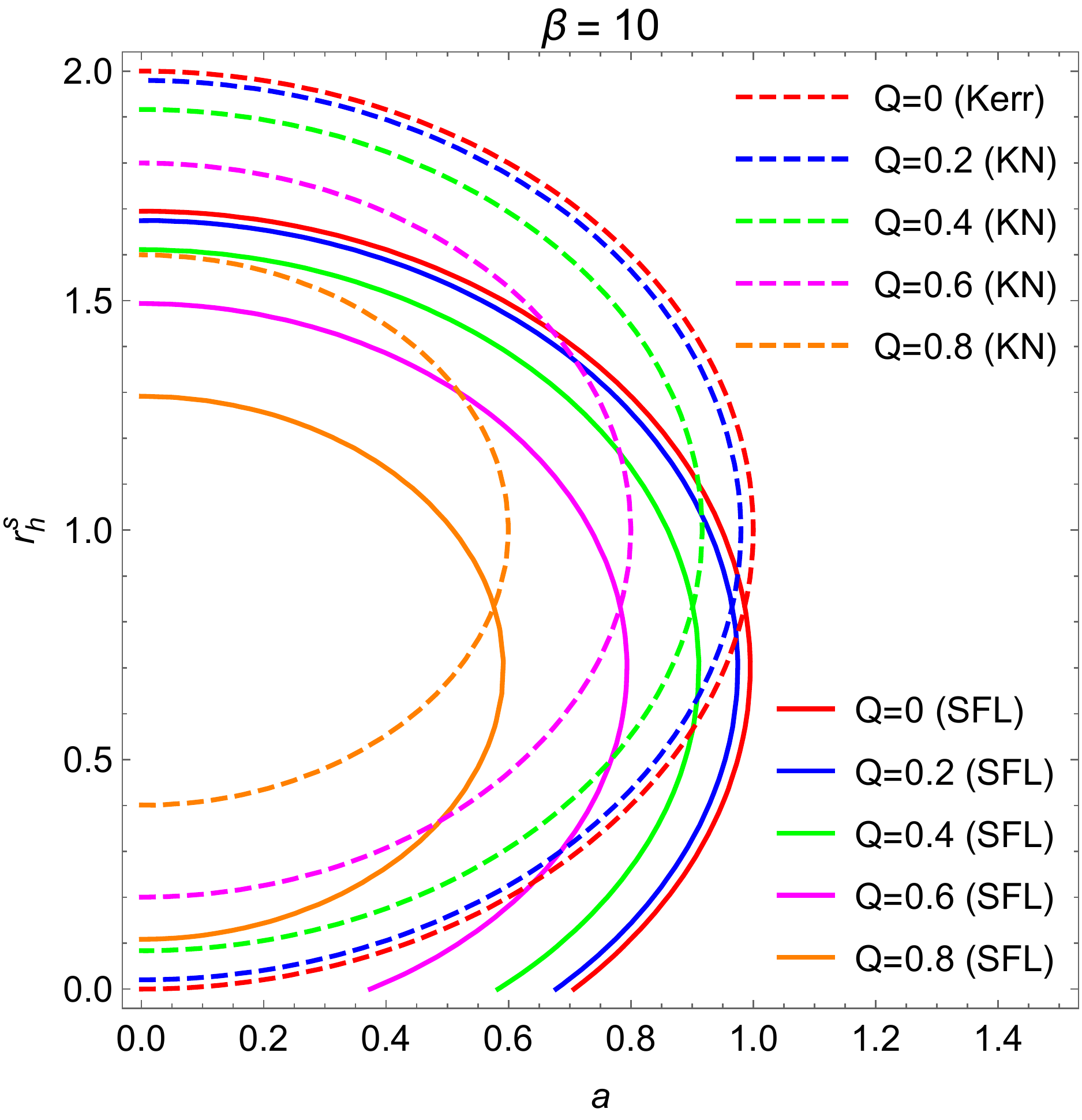}}
\subfigure{
\includegraphics[height=7.5cm,width=8cm]{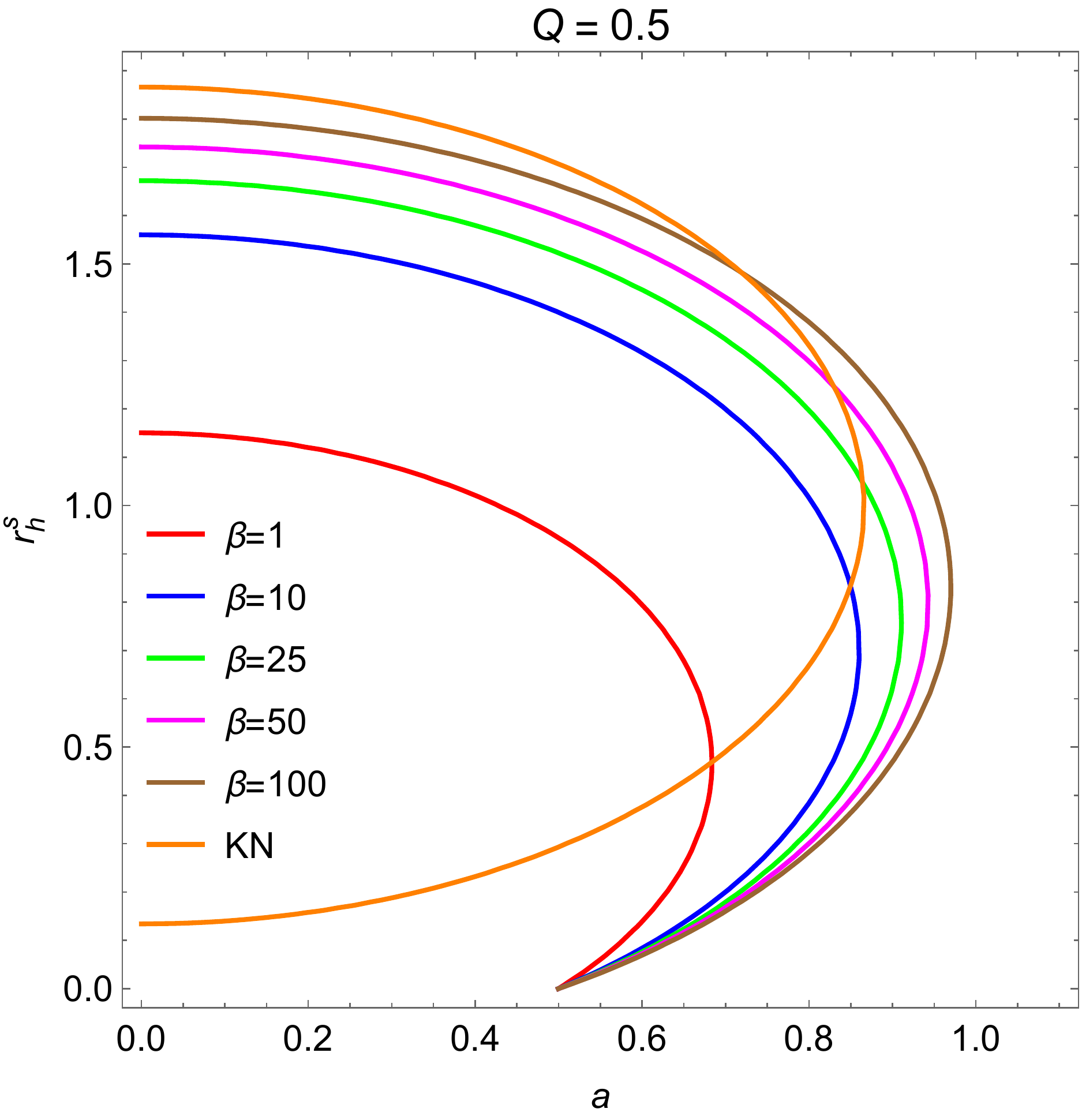}}
~~~~~~~
\subfigure{
\includegraphics[height=7.5cm,width=8cm]{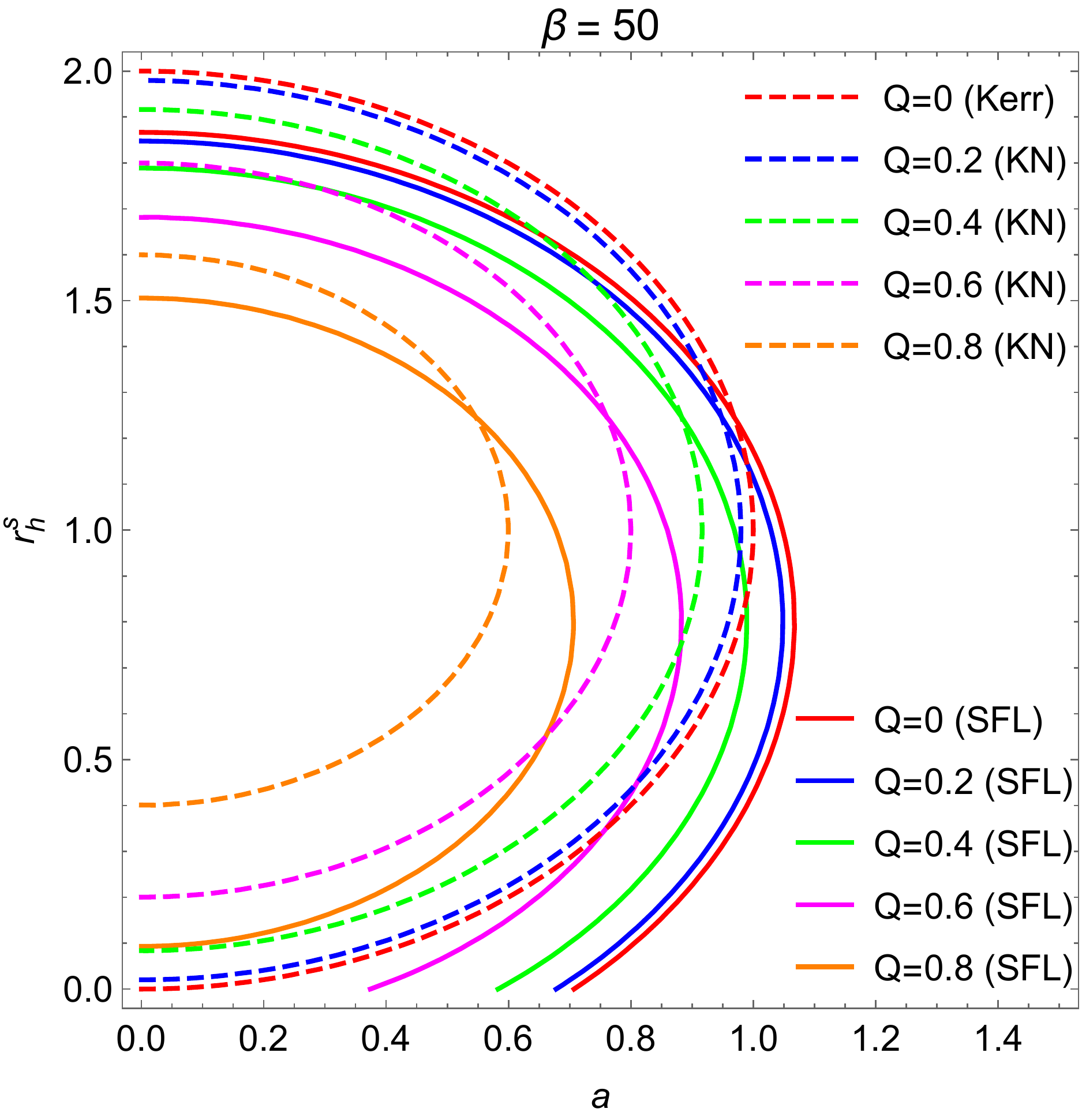}}
\subfigure{
\includegraphics[height=7.5cm,width=8cm]{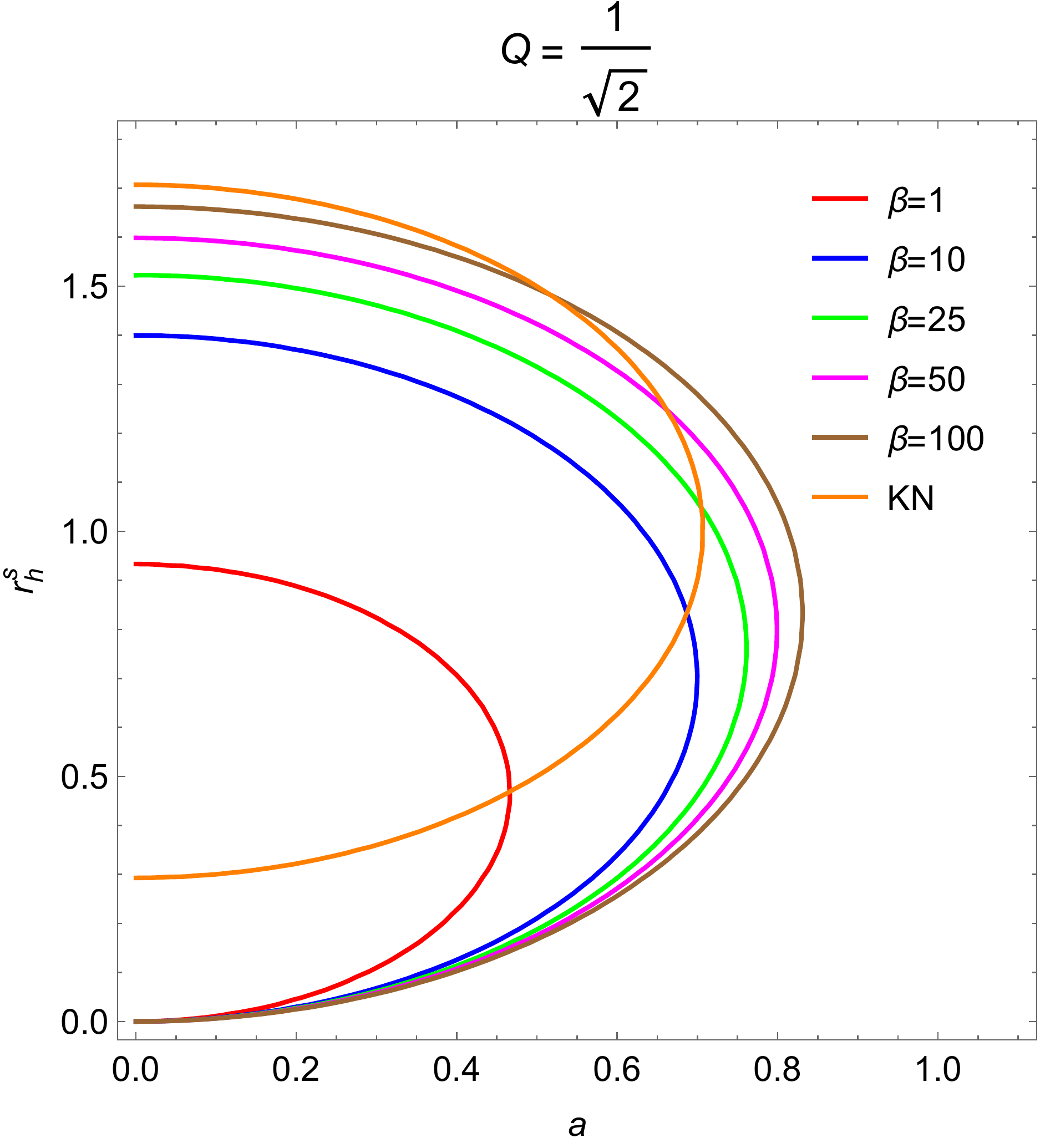}}
~~~~~~~
\subfigure{
\includegraphics[height=7.5cm,width=8cm]{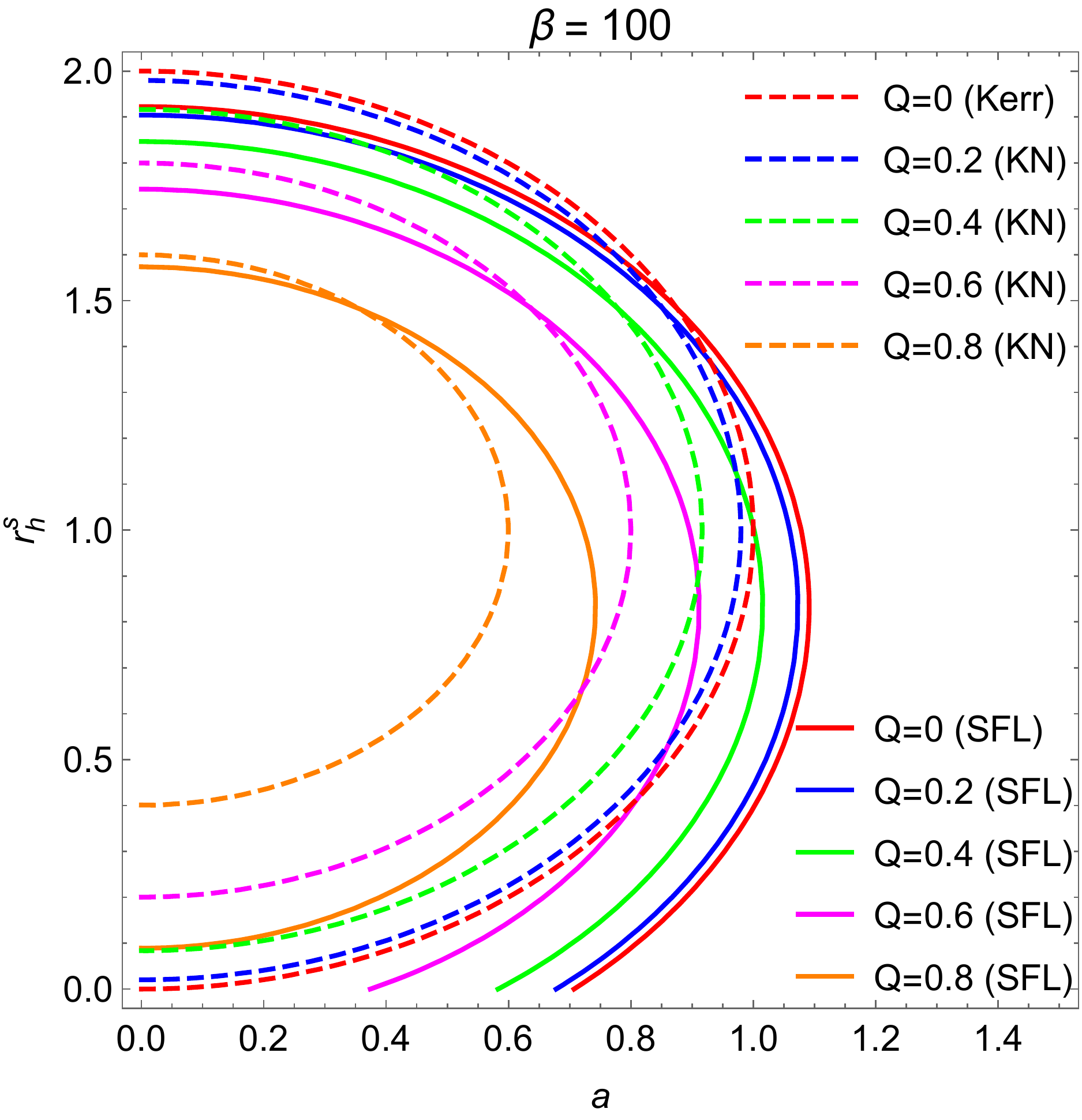}}
\end{center}
\caption{Plots for the behavior of horizon radius vs $a$ in strong field. The left panel corresponds to $Q=0$, $0.5$ and $1/\sqrt{2}$ for the curves corresponding to $\beta=1$, $10$, $25$, $50$ and $100$. The right panel shows the plots for $\beta=10$, $50$ and $100$ for the curves corresponding to $Q=0$, $0.2$, $0.4$, $0.6$ and $0.8$. KN stands for Kerr-Newman metric.}
\end{figure}

\section{Null Geodesics and Shadows}
To study the shadow and related observables for rotating metrics in strong field approximation, we need to derive geodesic equations by considering Hamilton-Jacobi formulation \cite{11,74}. The dynamics of bodies is widely studied by the Lagrangian and the Hamiltonian formalism in the context of Classical Mechanics, Electromagnetism, Quantum Mechanics etc. However, often it is difficult to deal with these formalisms in gravity theories. The usual procedures give two constants of motion, namely, the energy and the angular momentum along the axis of symmetry and the mass as a third constant. However, we require a fourth constant in order to solve the system of equations. The separability of Hamilton-Jacobi equation enables us to obtain a fourth constant called the Carter constant \cite{11,74}. Despite presenting the motivation for Hamilton-Jacobi formulation, we require the Lagrangian for deriving the relation between the constants of motion and the generalized momenta. The Lagrangian in the tensor notation is given as
\begin{equation}
\mathcal{L}=\frac{1}{2}g_{\mu\nu}\dot{x}^{\mu}\dot{x}^{\nu}. \label{39}
\end{equation}
The generalized momenta in the tensor form can be written as
\begin{equation}
p_\mu=g_{\mu\nu}\dot{x}^{\nu}. \label{40}
\end{equation}
Then the constants of motion corresponding to energy $E$ and the angular momentum $l$ along the axis of symmetry are related to the Lagrangian and the generalized momenta as
\begin{eqnarray}
E&=&-\frac{\partial\mathcal{L}}{\partial\dot{t}}=-g_{tt}\dot{t}-g_{\varphi t}\dot{\varphi}=p_t, \label{41} \\
l&=&\frac{\partial\mathcal{L}}{\partial\dot{\varphi}}=g_{\varphi t}\dot{t}+g_{\varphi \varphi}\dot{\varphi}=p_\varphi, \label{42}
\end{eqnarray}
where the dot represents the derivative w.r.t affine parameter $\tau$. The Eqs. (\ref{41}) and (\ref{42}) will be used in further calculations while dealing with the particular forms of generalized momenta and their relation with the constants of motion. In particular, these equations will be helpful in converting the system in terms of the constants $E$ and $l$. The Hamilton-Jacobi equation is given by
\begin{equation}
\frac{\partial \mathcal{S^J}}{\partial\tau}=-\frac{1}{2}g^{\mu\nu}\frac{\partial \mathcal{S^J}}{\partial x^\mu}\frac{\partial \mathcal{S^J}}{\partial x^\nu}. \label{43}
\end{equation}
Then the Jacobi action $\mathcal{S^J}$ can be chosen as
\begin{equation}
\mathcal{S^J}=\frac{1}{2}m\tau-Et+l\varphi+\mathcal{D}_r(r)+\mathcal{D}_\vartheta(\vartheta), \label{44}
\end{equation}
such that $m$ is the mass of the particle, $\mathcal{D}_r(r)$ and $\mathcal{D}_\vartheta(\vartheta)$ are unknown functions yet to be determined. Next, we use Eq. (\ref{44}) in Eq. (\ref{43}) and then apply the method of separation of variables as in \cite{11,74}, then by introducing the Carter constant $\mathcal{K}$, we obtain the geodesic equations for the photon $(m=0)$ given as
\begin{eqnarray}
\rho^2\dot{t}&=&a(l-aE\sin^2{\vartheta})+\frac{r^2+a^2}{\Delta_s}\bigg(E(r^2+a^2)-al\bigg), \label{45} \\
\rho^2\dot{r}&=&\pm\sqrt{\mathcal{R}}, \label{46} \\
\rho^2\dot{\vartheta}&=&\pm\sqrt{\Theta}, \label{47} \\
\rho^2\dot{\varphi}&=&(l\csc^2{\vartheta}-aE)+\frac{a}{\Delta_s}\bigg(E(r^2+a^2)-al\bigg), \label{48}
\end{eqnarray}
where
\begin{eqnarray}
\mathcal{R}(r)&=&\bigg(E(r^2+a^2)-al\bigg)^2-\Delta_s\mathcal{K}, \label{49} \\
\Theta(\vartheta)&=&\mathcal{K}-\frac{(aE\sin^2{\vartheta}-l)^2}{\sin^2{\vartheta}}. \label{50}
\end{eqnarray}

The geodesic equations govern the motion of the particles and other objects in the gravitational fields of massive bodies. Here, the particle under consideration is photon, so the above equations describe the motion and the trajectories followed by the massless photon. Since we are studying the motion of photons that in general, orbit the BH in circular geodesics so these photons stay on the surface of a sphere of radius $r=constant$. These orbits are characterized by the relations $\dot{r}=0$ and $\ddot{r}=0$ that implies $\mathcal{R}(r)=0$ and $\frac{d\mathcal{R}(r)}{dr}=0$. Then by introducing the abbreviations $L_E=\frac{l}{E}$ and $K_E=\frac{\mathcal{K}}{E^2}$, we get
\begin{eqnarray}
K_E&=&16r^2\frac{\Delta_s}{(\Delta'_s)^2}, \label{51} \\
aL_E&=&r^2+a^2-4r\frac{\Delta_s}{\Delta'_s}, \label{52}
\end{eqnarray}
where prime denotes the derivative w.r.t $r$. Since we know that all spherical null geodesics are unstable in the domain of outer communication, as in the case of Schwarzschild spacetime, at the observer position, the past-oriented null rays may asymptotically spiral towards the circular null geodesics, which behave as the limit curves. So the boundary of the shadow can be determined by the photon region. We consider two angles; a colatitude angle $\theta$ and an azimuthal angle $\psi$ in the observer's sky. These angles determine the initial direction of all null rays into the past emerging from the observer position. We can define these angles with respect to the orthonormal tetrad
\begin{eqnarray}
e_0&=&\frac{(r^2+a^2)\partial_t + a\partial_\varphi}{\rho\sqrt{\Delta_s}}\bigg|_{(r_0,\vartheta_0)}, \label{53} \\
e_1&=&\frac{\partial_\vartheta}{\rho}\bigg|_{(r_0,\vartheta_0)}, \label{54} \\
e_2&=&\frac{-\partial_\varphi - a\sin^2{\vartheta}\partial_t}{\rho\sin\vartheta}\bigg|_{(r_0,\vartheta_0)}, \label{55} \\
e_3&=&-\frac{\sqrt{\Delta_s}\partial_r}{\rho}\bigg|_{(r_0,\vartheta_0)}, \label{56}
\end{eqnarray}
for an observer at $(r_0,\vartheta_0)$. For the geometrical configuration see Fig. $\textbf{7}$ and $\textbf{8}$ in \cite{18}. It can be easily verified that $e_i$ are orthonormal for the metric corresponding to $\Delta_s$. The timelike vector $e_0$ is considered to be the four velocity of the observer. Meanwhile $e_3$ is pointing towards the center of the BH. Let $\Upsilon(\varrho)$ be a null ray with coordinates $(r(\varrho),\vartheta(\varrho),\varphi(\varrho),t(\varrho))$, then the tangent to this curve can be written in two ways:
\begin{eqnarray}
\dot{\Upsilon}&=&\dot{r}\partial_r+\dot{\vartheta}\partial_\vartheta+\dot{\varphi}\partial_\varphi+\dot{t}\partial_t, \label{57} \\
\dot{\Upsilon}&=&\xi(-e_0+\sin\theta\cos\psi e_1+\sin\theta\sin\psi e_2+\cos\theta e_3). \label{58}
\end{eqnarray}
We have defined the celestial coordinates $\theta$ and $\psi$ for the light ray in the observer's sky. The scale factor $\xi$ is obtained using Eqs. (\ref{41}) and (\ref{42}) as
\begin{eqnarray}
\xi=g(\dot{\Upsilon},e_0)=\frac{al-(r^2+a^2)E}{\rho\sqrt{\Delta_s}}\bigg|_{(r_0,\vartheta_0)}. \label{59}
\end{eqnarray}
By comparing the coefficients of $\partial_r$ and $\partial_\varphi$ in Eqs. (\ref{57}) and (\ref{58}), we get
\begin{eqnarray}
\sin\psi&=&\frac{\sin\vartheta}{\sqrt{\Delta_s}\sin\theta}\bigg(\frac{\rho^2\Delta_s\dot{\varphi}}{(r^2+a^2)E-al}-a\bigg)\bigg|_{(r_0,\vartheta_0)}, \label{60} \\
\cos\theta&=&\frac{\rho^2\dot{r}}{(r^2+a^2)E-al}\bigg|_{(r_0,\vartheta_0)}. \label{61}
\end{eqnarray}
By inserting the values of $\dot{r}$ and $\dot{\varphi}$ from Eqs. (\ref{46}) and (\ref{48}), we get
\begin{eqnarray}
\sin\psi(r_p^s)&=&\frac{L_E(r_p^s)-a\sin^2\vartheta_0}{\sqrt{K_E(r_p^s)}\sin\vartheta_0}, \label{62} \\
\sin\theta(r_p^s)&=&\frac{\sqrt{K_E(r_p^s)}\sqrt{\Delta_s(r_0)}}{r_0^2+a^2-aL_E(r_p^s)}. \label{63}
\end{eqnarray}

The quantities $K_E(r_p^s)$ and $L_E(r_p^s)$ are defined by Eqs. (\ref{51}) and (\ref{52}). Since the light rays spiral asymptotically towards the spherical photon region with radius $r_p^s$, then the governing equations depend upon $r_p^s$ because the photon region has a maximum and minimum radius such that $r_p^s$ has all possible values in the interval $[r^s_{p,min},r^s_{p,max}]$. The extremal values of $r_p^s$ are determined by the relations $\sin\psi(r_p^s)=1$ for $r^s_{p,min}$ and $\sin\psi(r_p^s)=-1$ for $r^s_{p,max}$. For the static case, there is a photon sphere which has a fixed radius and as a result we can not consider $r_p^s$ as a parameter. This value of $r_p^s$ gives a unique value of $K_E(r_p^s)$ but $L_E(r_p^s)$ can not be calculated. Hence, we can consider $L_E(r_p^s)$ as the parameter, whose values exist in the interval determined by $\Theta(\vartheta_0)=0$.

\subsection{Shadows Observed by a Nearer Observer}
Now we will examine the shadows observed by an observer in the vicinity of the BH. The Eqs. (\ref{51}) and (\ref{52}) are further written by using the definition of metric functions as
\begin{eqnarray}
K_E^s(r_p^s)&=&\frac{4(r_p^s)^2\bigg[(r_p^s)^2-2Mr_p^s+a^2+Q^2-\frac{C^2\kappa^2}{2}+\frac{16C^{3/2}\kappa^2r_p^s}{15\beta^{1/4}}\bigg]}{\bigg(r_p^s-M+\frac{8C^{3/2}\kappa^2}{15\beta^{1/4}}\bigg)^2}, \label{64} \\
aL_E^s(r_p^s)&=&\frac{(r_p^s)^3-3\bigg(M-\frac{8C^{3/2}\kappa^2}{15\beta^{1/4}}\bigg)(r_p^s)^2+(a^2+2Q^2-C^2\kappa^2)r_p^s+\bigg(M-\frac{8C^{3/2}\kappa^2}{15\beta^{1/4}}\bigg)a^2}{M-r_p^s-\frac{8C^{3/2}\kappa^2}{15\beta^{1/4}}}. \label{65}
\end{eqnarray}
Thus, the equations for celestial coordinates in the observer's sky become
\begin{eqnarray}
\sin\psi_s(r_p^s)&=&\frac{-1}{2ar_p^s\sin{\vartheta_0}\bigg[(r_p^s)^2-2Mr_p^s+a^2+Q^2-\frac{C^2\kappa^2}{2}+\frac{16C^{3/2}\kappa^2r_p^s}{15\beta^{1/4}}\bigg]^{1/2}}\bigg[(r_p^s)^3-3\bigg(M-\frac{8C^{3/2}\kappa^2}{15\beta^{1/4}}\bigg)(r_p^s)^2 \nonumber \\
&&+(a^2+2Q^2-C^2\kappa^2)r_p^s+\bigg(M-\frac{8C^{3/2}\kappa^2}{15\beta^{1/4}}\bigg)a^2-a^2\sin^2{\vartheta_0}\bigg(M-r_p^s-\frac{8C^{3/2}\kappa^2}{15\beta^{1/4}}\bigg)\bigg], \label{66} \\
\sin\theta_s(r_p^s)&=&\frac{2r_p^s}{(r_p^s)^3-3M(r_p^s)^2-Mr_0^2+(r_0^2+2a^2+2Q^2-C^2\kappa^2)r_p^s+\frac{8C^{3/2}\kappa^2}{15\beta^{1/4}}\bigg(r_0^2+3(r_p^s)^2\bigg)}\bigg[(r_p^s)^2-2Mr_p^s \nonumber \\
&&+a^2+Q^2-\frac{C^2\kappa^2}{2}+\frac{16C^{3/2}\kappa^2r_p^s}{15\beta^{1/4}}\bigg]^{1/2}\bigg[r_0^2-2Mr_0+a^2+Q^2-\frac{C^2\kappa^2}{2}+\frac{16C^{3/2}\kappa^2r_0}{15\beta^{1/4}}\bigg]^{1/2}. \label{67}
\end{eqnarray}
For a stereographic projection, we have the Cartesian coordinates
\begin{eqnarray}
x(r_p^s)&=&-2\tan\bigg(\frac{\theta_s(r_p^s)}{2}\bigg)\sin\psi_s(r_p^s), \label{68} \\
y(r_p^s)&=&-2\tan\bigg(\frac{\theta_s(r_p^s)}{2}\bigg)\cos\psi_s(r_p^s). \label{69}
\end{eqnarray}

The shadows for an equatorial observer in the strong field region are plotted in the Fig. $\textbf{3}$. The top panel shows the boundary curves for different values of spin. It is found that with increasing value of $a$, the shadows are shifted in the positive x-direction. Moreover, with increasing charge, the size of the shadows is reduced. In the middle panel we see that the shadow size is decreased with increasing charge as expected. However, by increasing $\beta$, the shadow size increases and the flatness on one side due to spin decreases under the effect of $\beta$ and other parameters. The first plot in the bottom panel corresponds to the case without spin. It can be seen that each curve has equal magnitude of $y$,$-y$,$x$ and $-x$-intercepts which means that the shadows are exactly circular. Also, the overall size of shadow is increased by increase in $\beta$. Almost the same behavior can be seen for other two plots apart from the effect of spin. The inclusion of spin shifted the shadows rightwards and some flatness is observed with increasing $a$. In Fig. $\textbf{3}$, the shadows for a fixed observer at $r_0=5$ are recorded. However, in Fig. $\textbf{4}$, the shadows for an equatorial observer at different positions are plotted. The image of the shadow became smaller as the observer moved away from the BH. Also the spin shifted the image rightwards and created flatness but the nonlinear electrodynamics diminished the flatness simultaneously. Note that the shadows for Kerr and Kerr-Newman exhibit a visible flatness on one side of the shadow. However, in our case, although the shadows are elongated due to spin but the flatness is diminished by the effect of $\beta$, $C$ and $\kappa$. Also, when $a=0$, the size of shadow is different for the standard Schwarzschild and Reissner-Nordstr\"{o}m metric as compared to the strong field metric under consideration.
\begin{figure}[t]
\begin{center}
\subfigure{
\includegraphics[height=4.93cm,width=5.6cm]{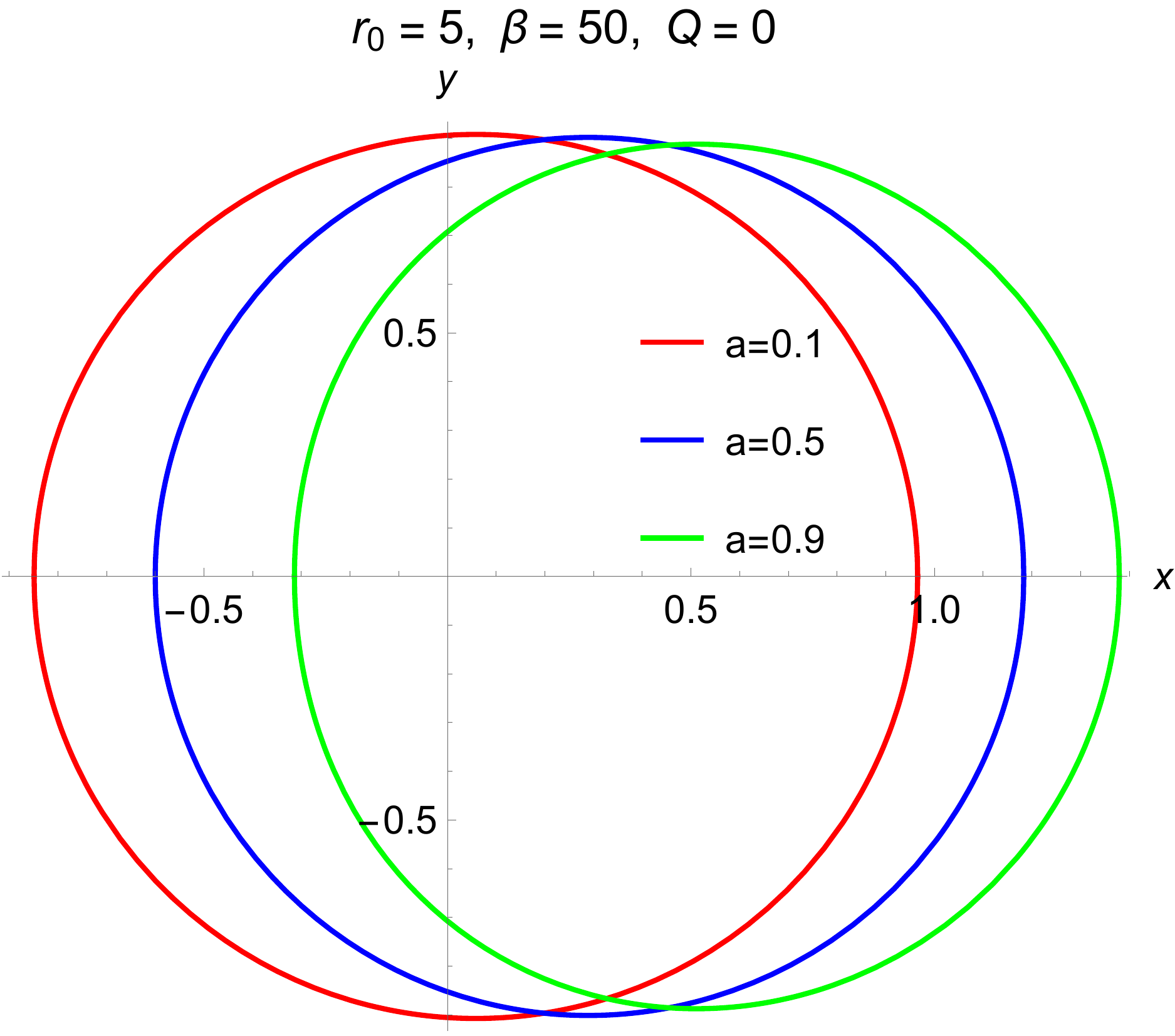}}
~
\subfigure{
\includegraphics[height=4.79cm,width=5.6cm]{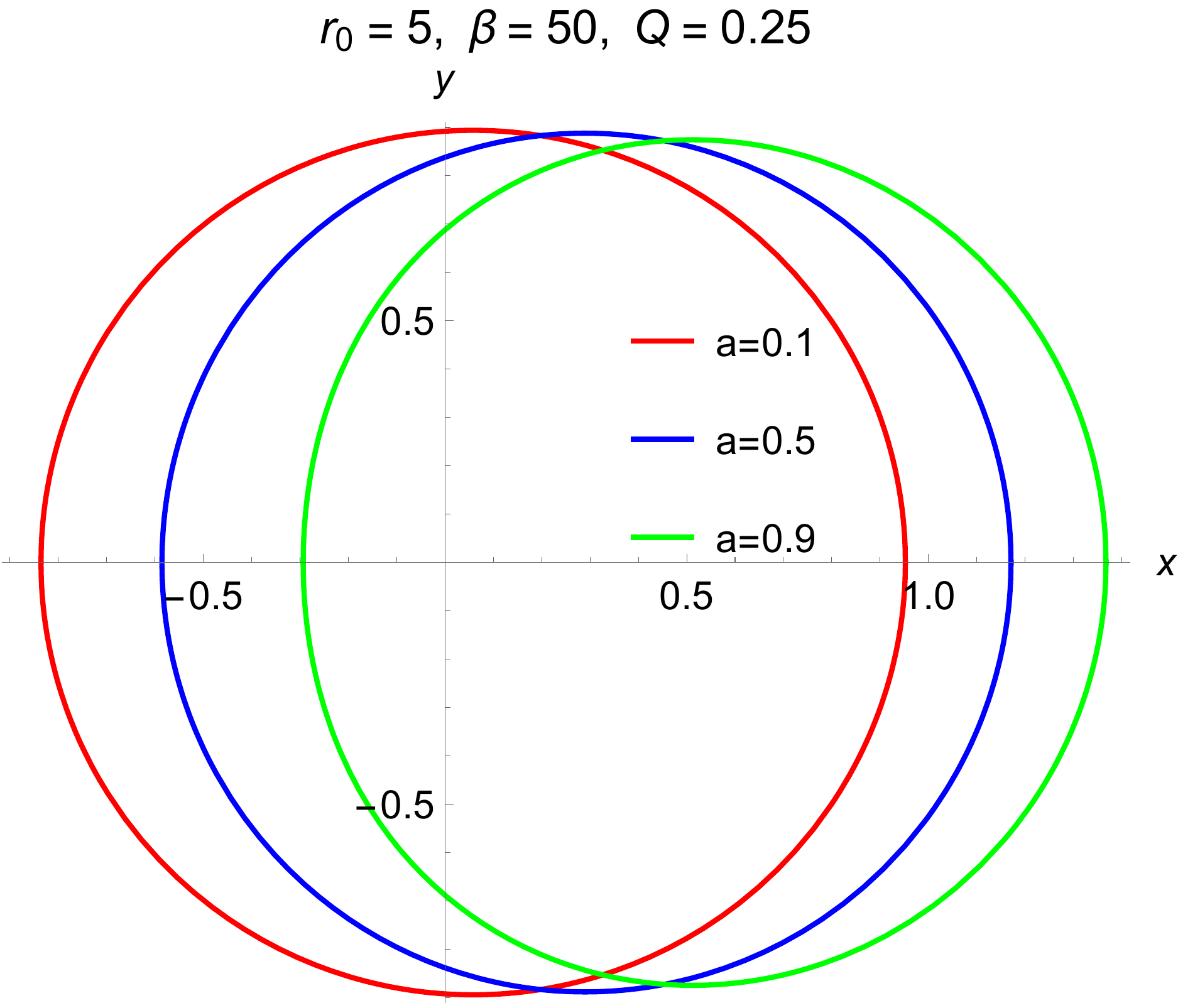}}
~
\subfigure{
\includegraphics[height=5.09cm,width=5.6cm]{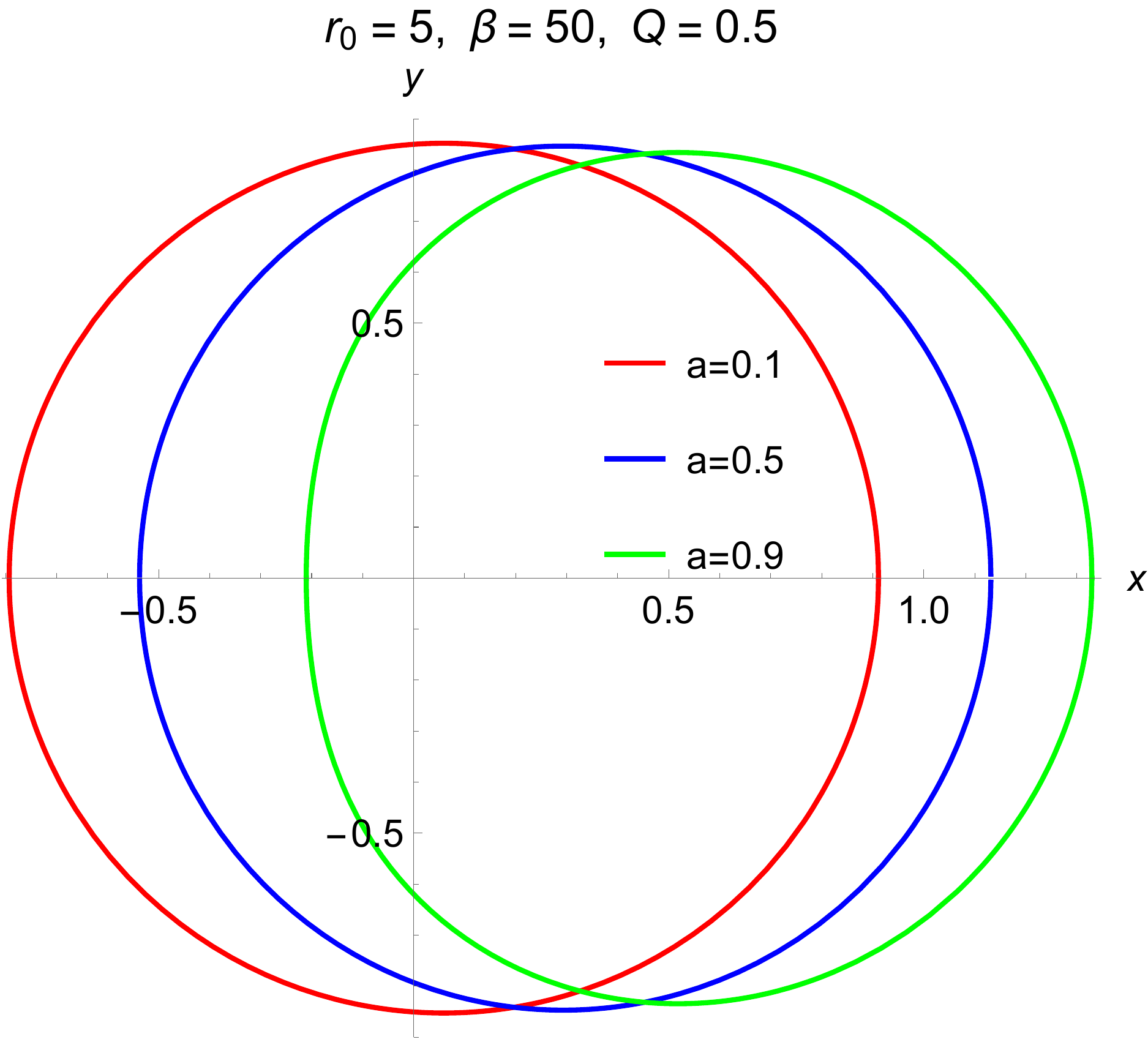}}
\subfigure{
\includegraphics[height=5.79cm,width=5.6cm]{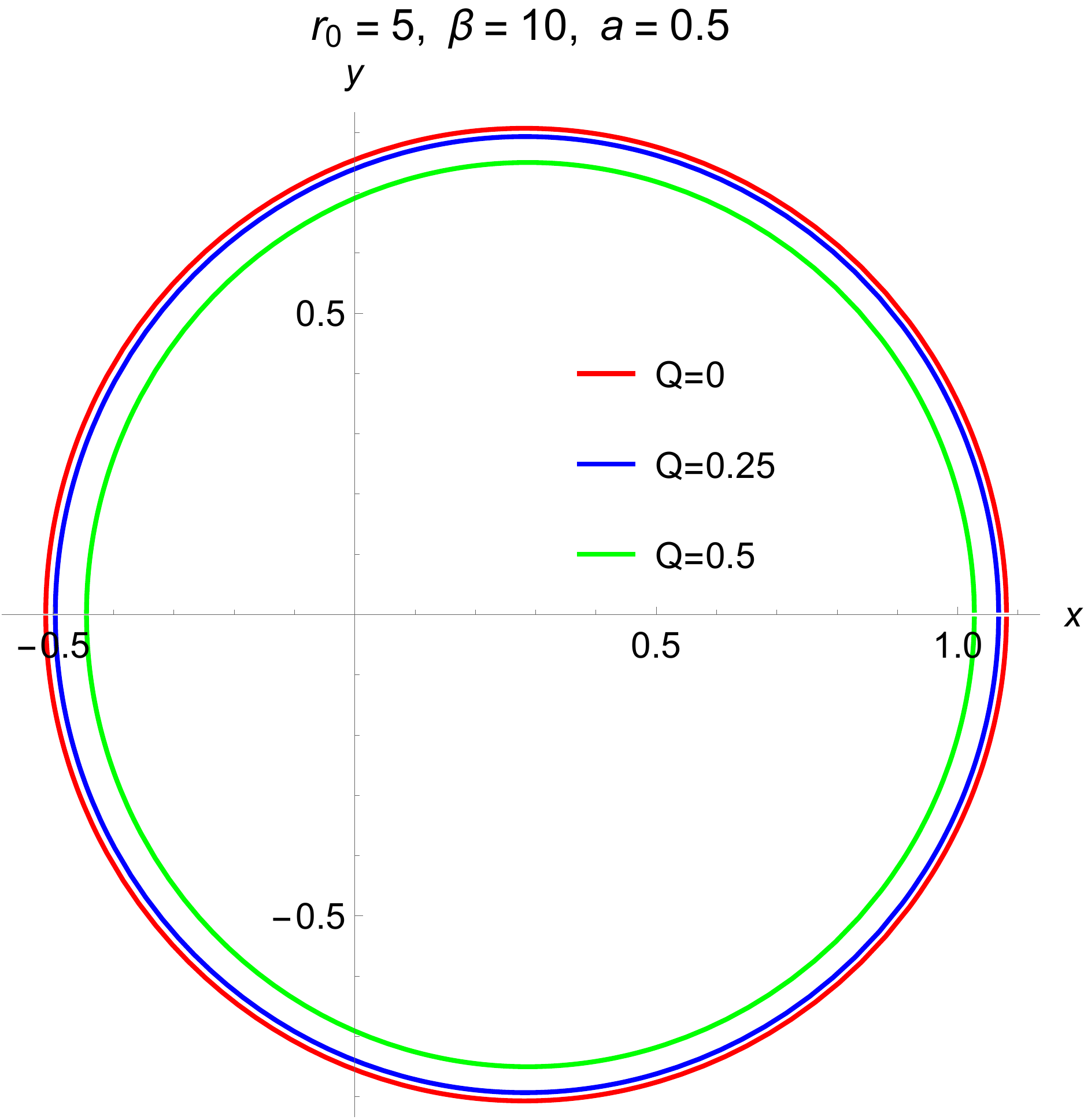}}
~
\subfigure{
\includegraphics[height=5.96cm,width=5.6cm]{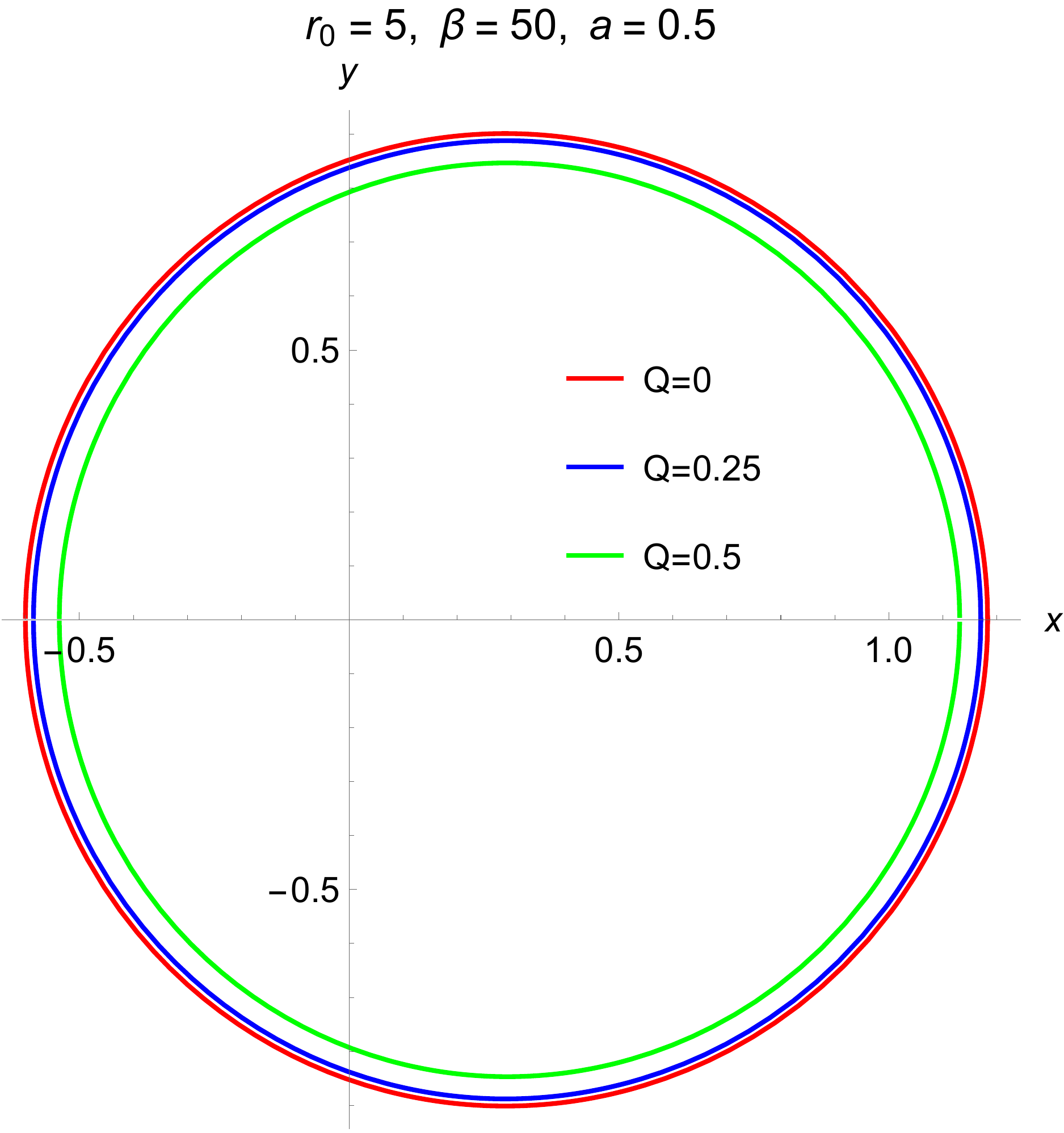}}
~
\subfigure{
\includegraphics[height=5.87cm,width=5.6cm]{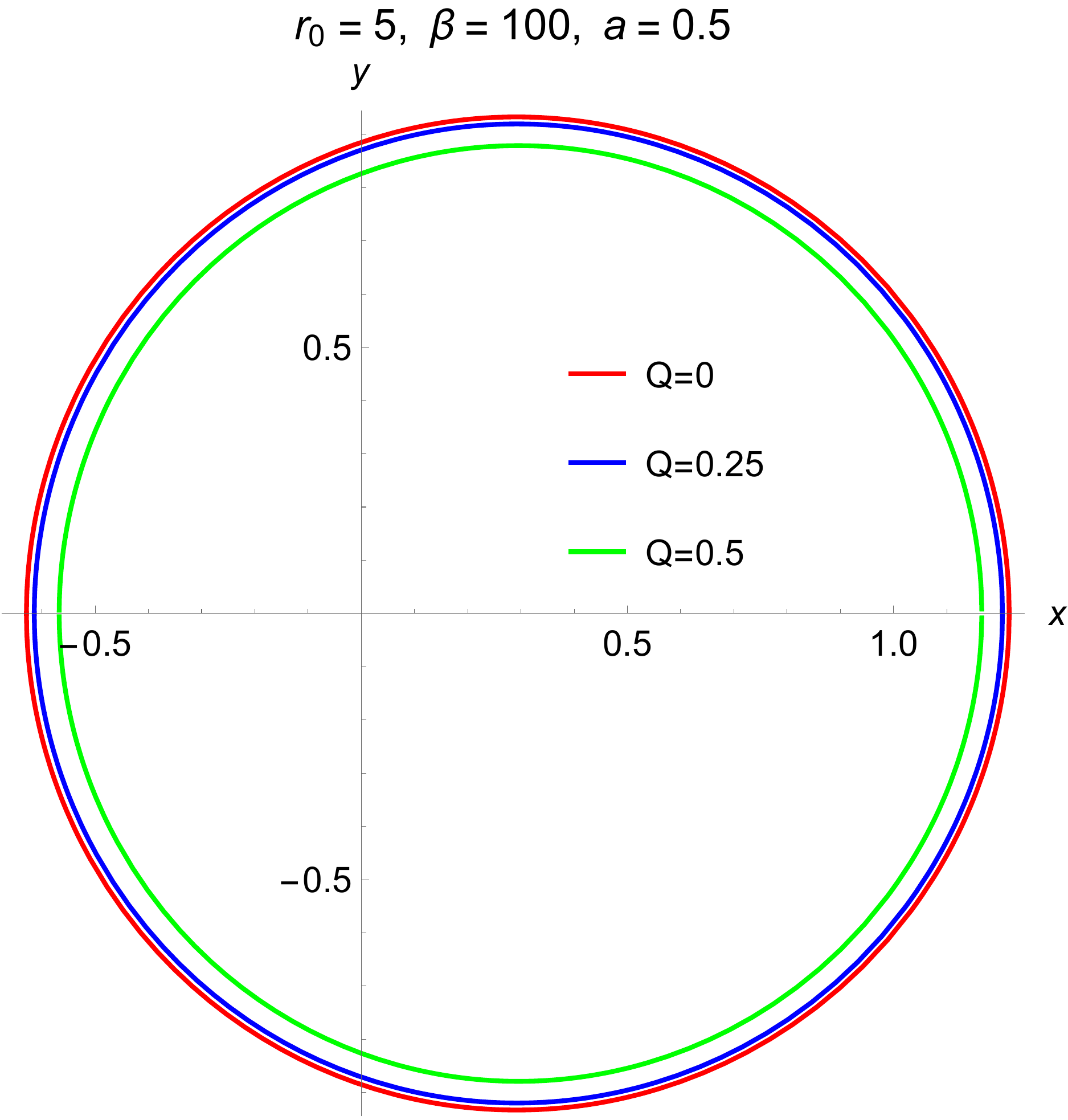}}
\subfigure{
\includegraphics[height=5.88cm,width=5.6cm]{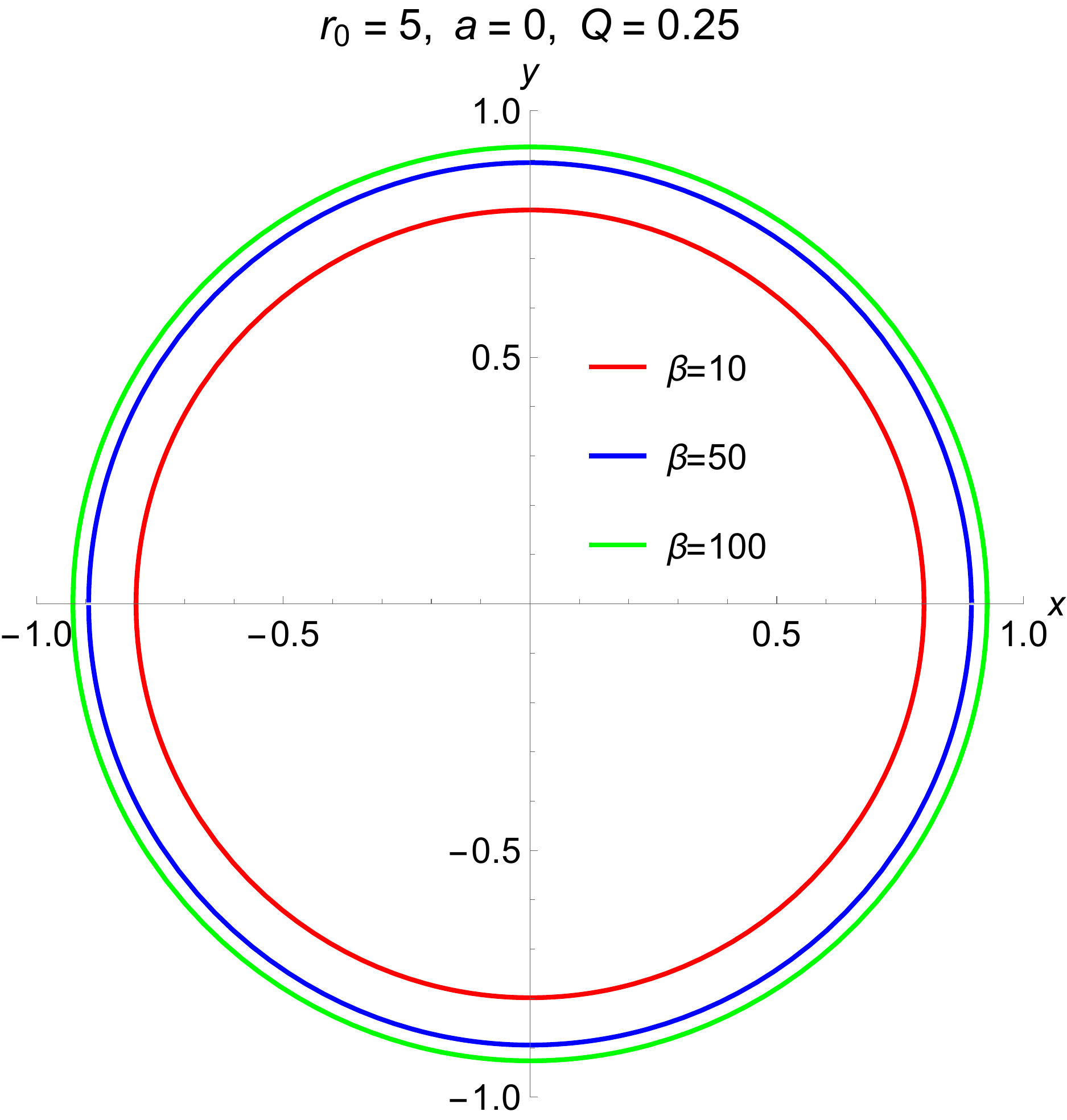}}
~
\subfigure{
\includegraphics[height=5.84cm,width=5.6cm]{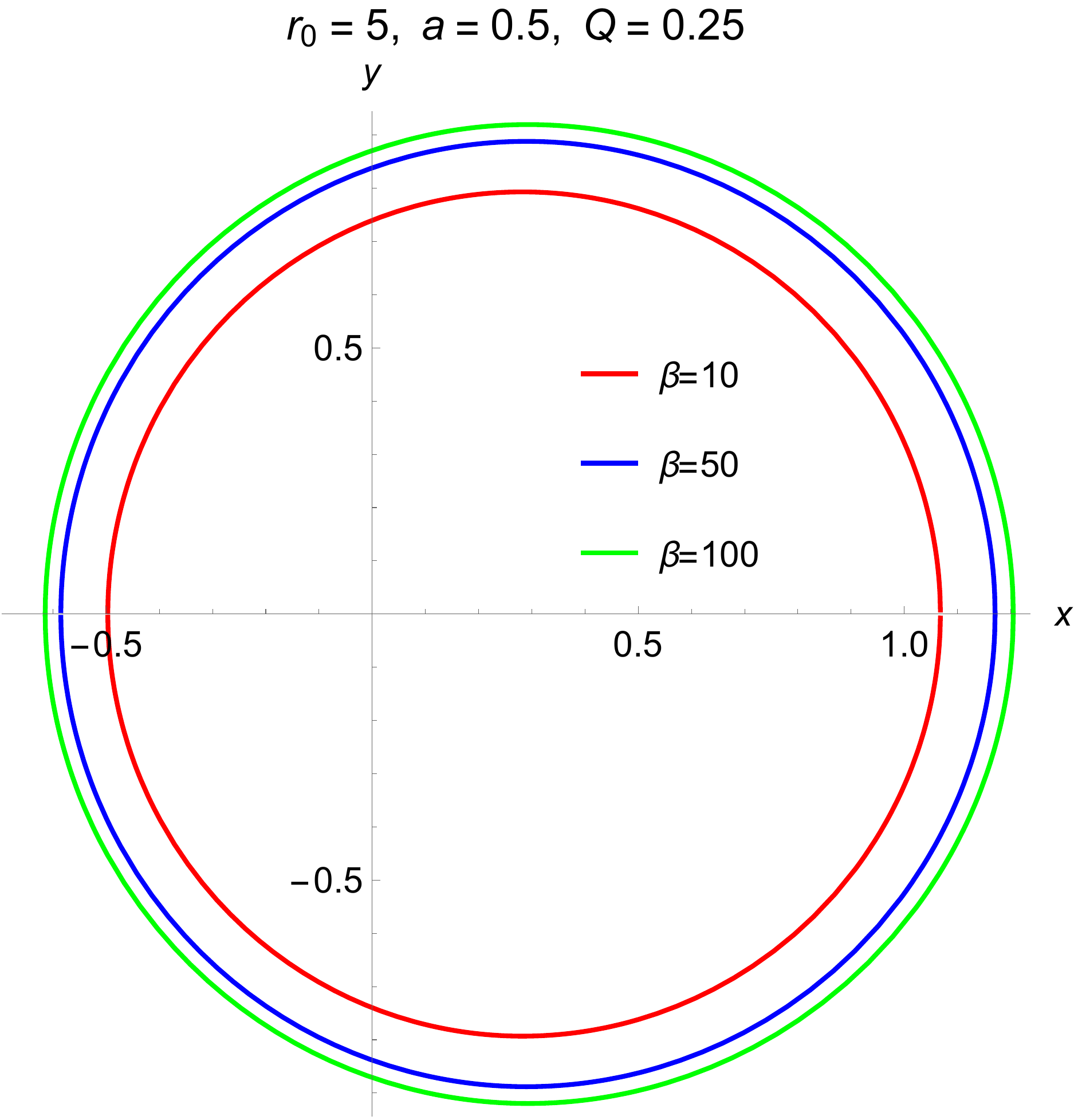}}
~
\subfigure{
\includegraphics[height=5.96cm,width=5.6cm]{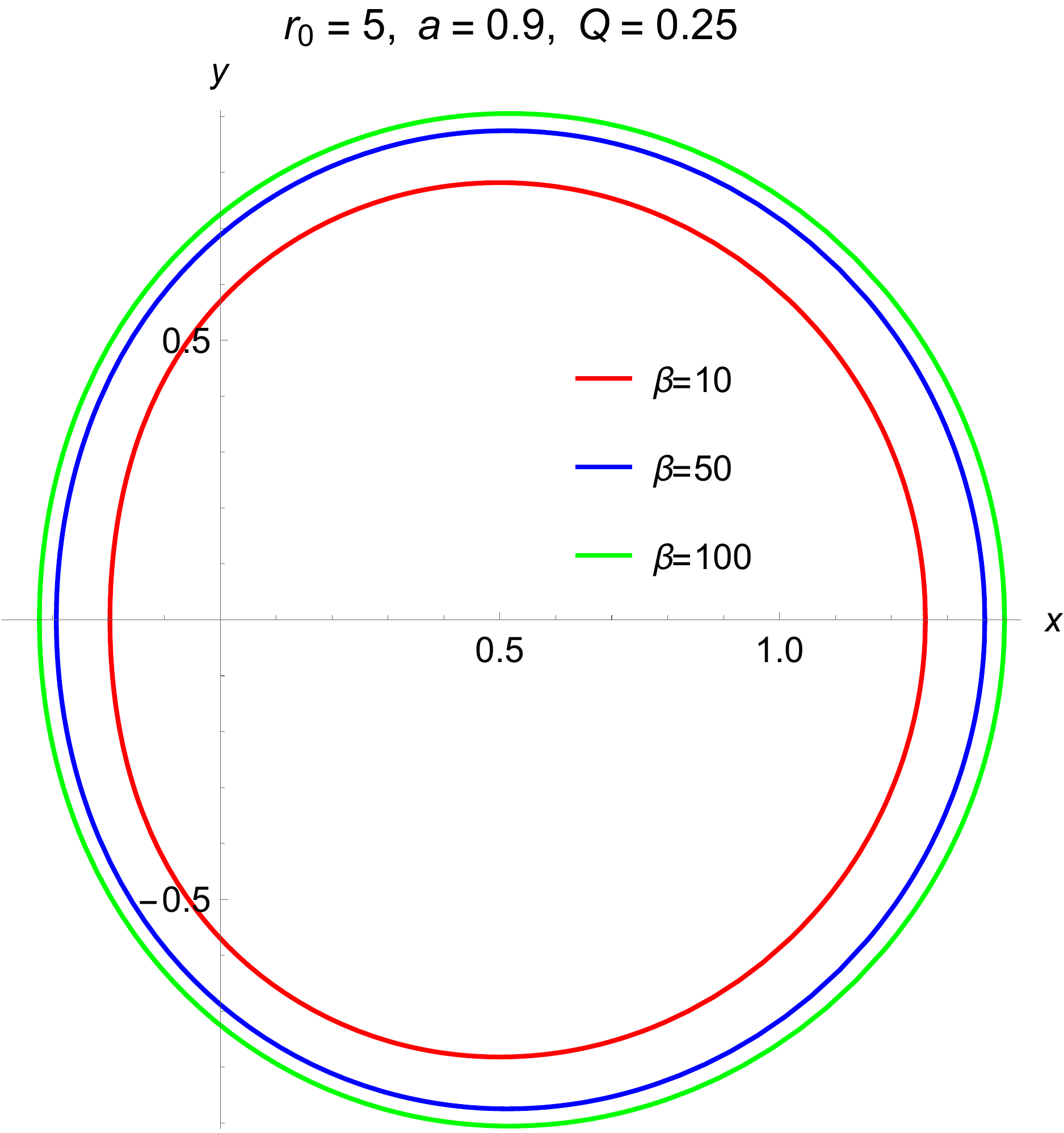}}
\end{center}
\caption{Stereographic projection of shadows in strong field w.r.t an equatorial observer at $r_0=5$. The top panel corresponds to the curves for $a=0.1$, $0.5$ and $0.9$. The middle panel corresponds to the curves for $Q=0$, $0.25$ and $0.5$. The bottom panel corresponds to the curves for $\beta=10$, $50$ and $100$.}
\end{figure}

\begin{figure}[t]
\begin{center}
\includegraphics[height=7cm,width=6.93cm]{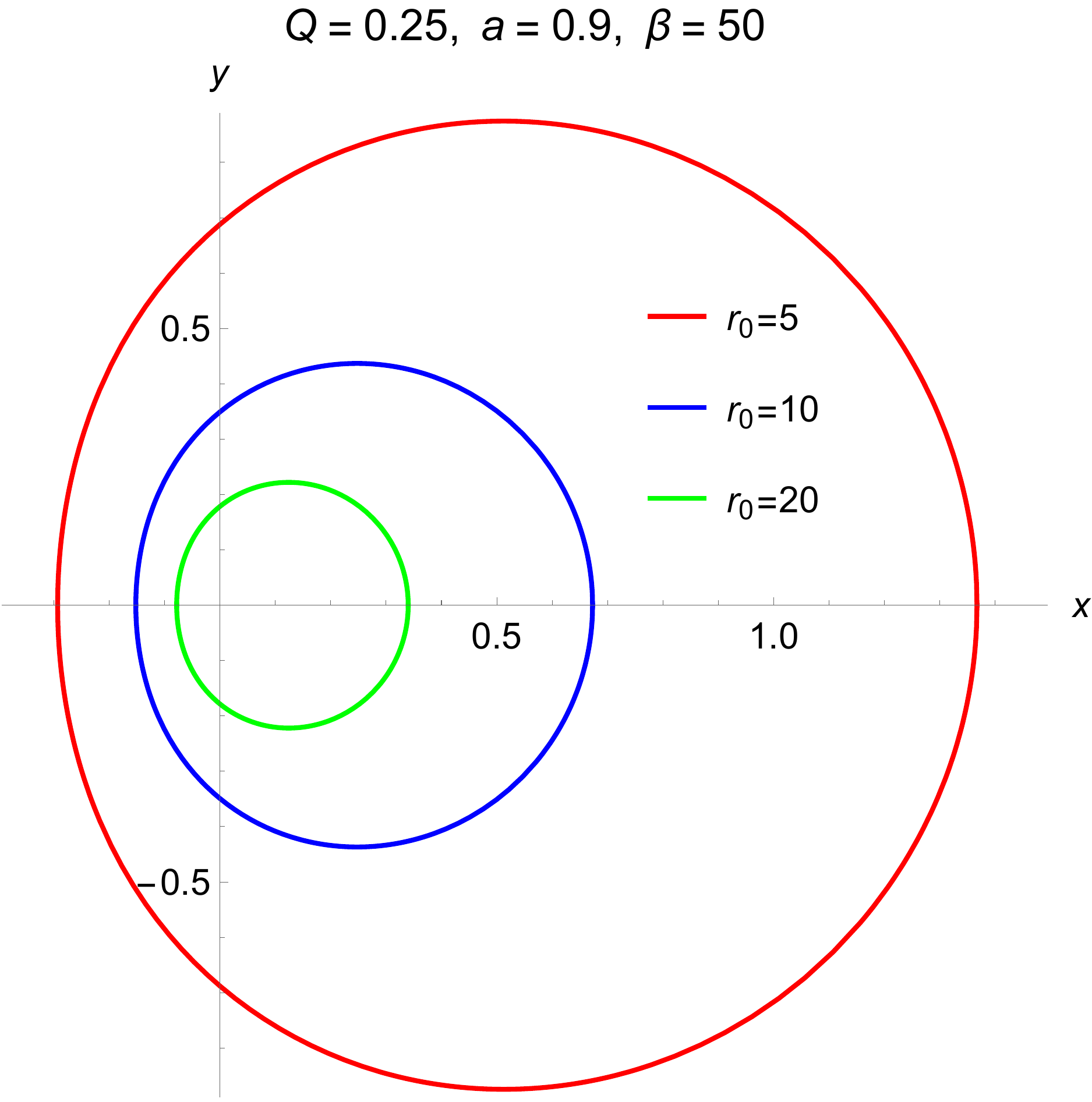}
\end{center}
\caption{Behavior of shadows in strong field w.r.t an equatorial observer at $r_0=5$, $10$ and $20$ for rotating case with fixed $a$, $Q$ and $\beta$.}
\end{figure}

\subsection{Shadows Observed by a Remote Observer}
The Bardeen's procedure for calculating the shadows is based on two impact parameters and the assumption that the observer is at a large distance from the BH. Within the framework of the method given in this study, we can assume a remote observer that is independent of the choice of the spacetime \cite{10}. It means that, the photon sphere formed by the trapping of photons in the vicinity of the BH and giving rise to the shadow, is observed by a remote observer. In other words, the observer looks directly into the BH and its surroundings where the photon sphere exists. For a remote observer, the linearization implies that $\frac{M}{r_0}\rightarrow0$, which gives
\begin{eqnarray}
\theta_s(r_p^s)&=&\frac{2r_p^s\bigg[(r_p^s)^2-2Mr_p^s+a^2+Q^2-\frac{C^2\kappa^2}{2}+\frac{16C^{3/2}\kappa^2r_p^s}{15\beta^{1/4}}\bigg]^{1/2}}{r_0\bigg(r_p^s-M+\frac{8C^{3/2}\kappa^2}{15\beta^{1/4}}\bigg)}. \label{70}
\end{eqnarray}
Then the following stereographic projection coordinates are obtained:
\begin{eqnarray}
x(r_p^s)&=&\frac{a\sin^2\vartheta_0-L_E^s(r_p^s)}{r_0\sin\vartheta_0}, \label{71} \\
y(r_p^s)&=&\pm\frac{1}{r_0}\sqrt{K_E^s(r_p^s)-\frac{(L_E^s(r_p^s)-a\sin^2\vartheta_0)^2}{\sin^2\vartheta_0}}, \label{72}
\end{eqnarray}
where the Eqs. (\ref{64}), (\ref{65}) and (\ref{66}) remain the same in this case because these are independent of the position of the observer. The behavior of shadow images perceived by a remote observer can be seen in Fig. $\textbf{5}$. In the top panel, the spin caused the shift of images towards right and not an exact flatness is observed which is diminished by the effect of $\beta$, $C$ and $\kappa$. As one goes from left to right in the panel, due to the increase in charge, the shadow size is decreased. In the middle panel, very small differences are measured in the shadows. The image size reduced with the increasing charge and with increasing $\beta$ in the panel, the image size is increased. The first plot in the bottom panel corresponds to $a=0$ and the shadow images can be identified as perfect circles. However, with the increase in $\beta$ in each plot, the image size is also increased. The spin increases from left to right in the panel, with this increase in spin, the shadow images move rightwards with an elongation in the shadow and a possible flatness is again diminished by the presence of $\beta$, $C$ and $\kappa$. The difference in the curves on $-x$-axis in the last plot in the panel shows a very prominent effect of nonlinearity of electrodynamics. It can be verified that the shadow size for the Schwarzschild and Reissner-Nordstr\"{o}m BHs is different from the shadow size in this study for the case $a=0$. Moreover, the Kerr and Kerr-Newman BHs exhibit a clear flatness on one side of the shadow, however, in our study, the flatness is diminished. These effects and differences of images will be studied in the following subsection in terms of distortion.
\begin{figure}[t]
\begin{center}
\subfigure{
\includegraphics[height=4.93cm,width=5.6cm]{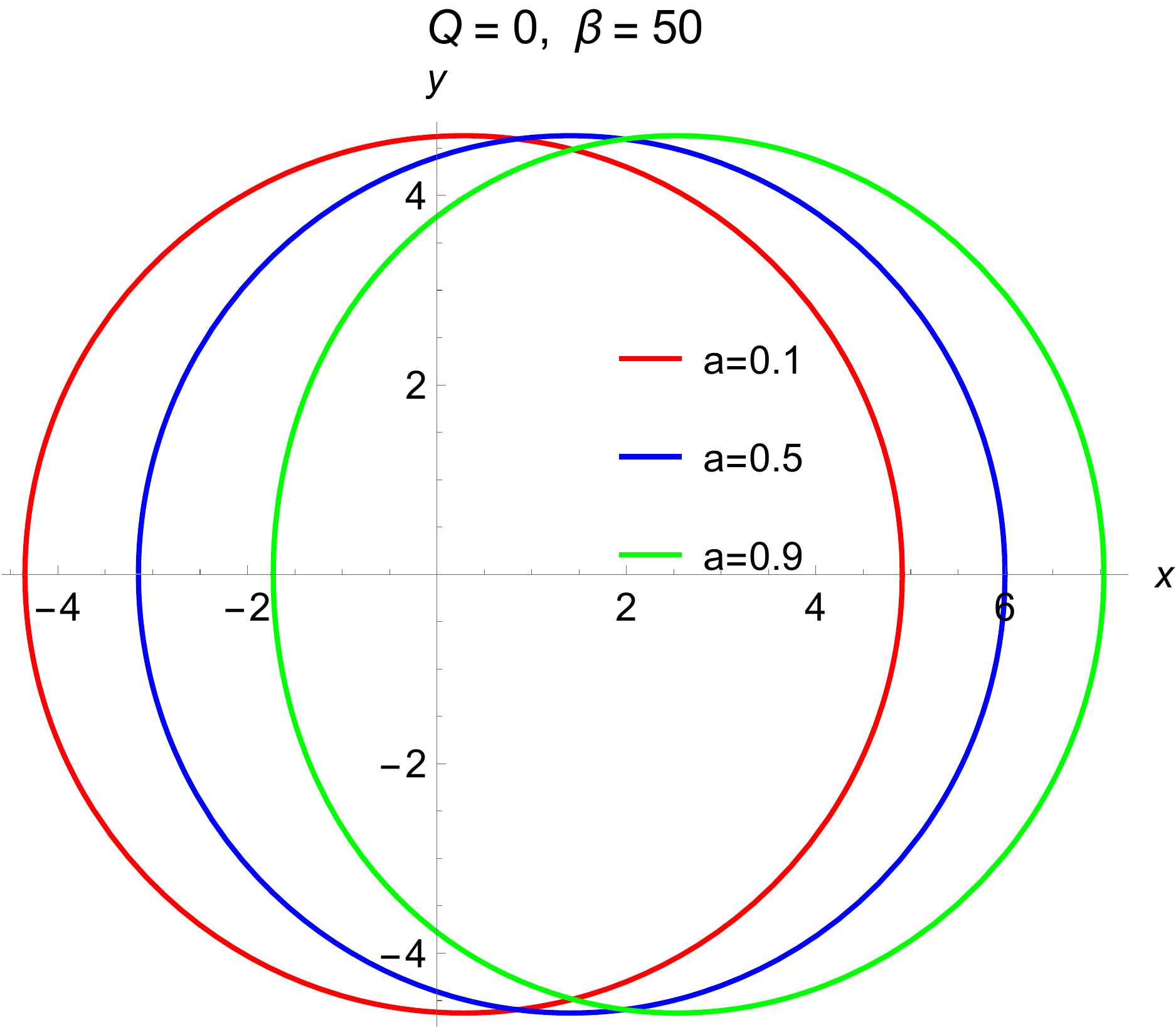}}
~
\subfigure{
\includegraphics[height=4.79cm,width=5.6cm]{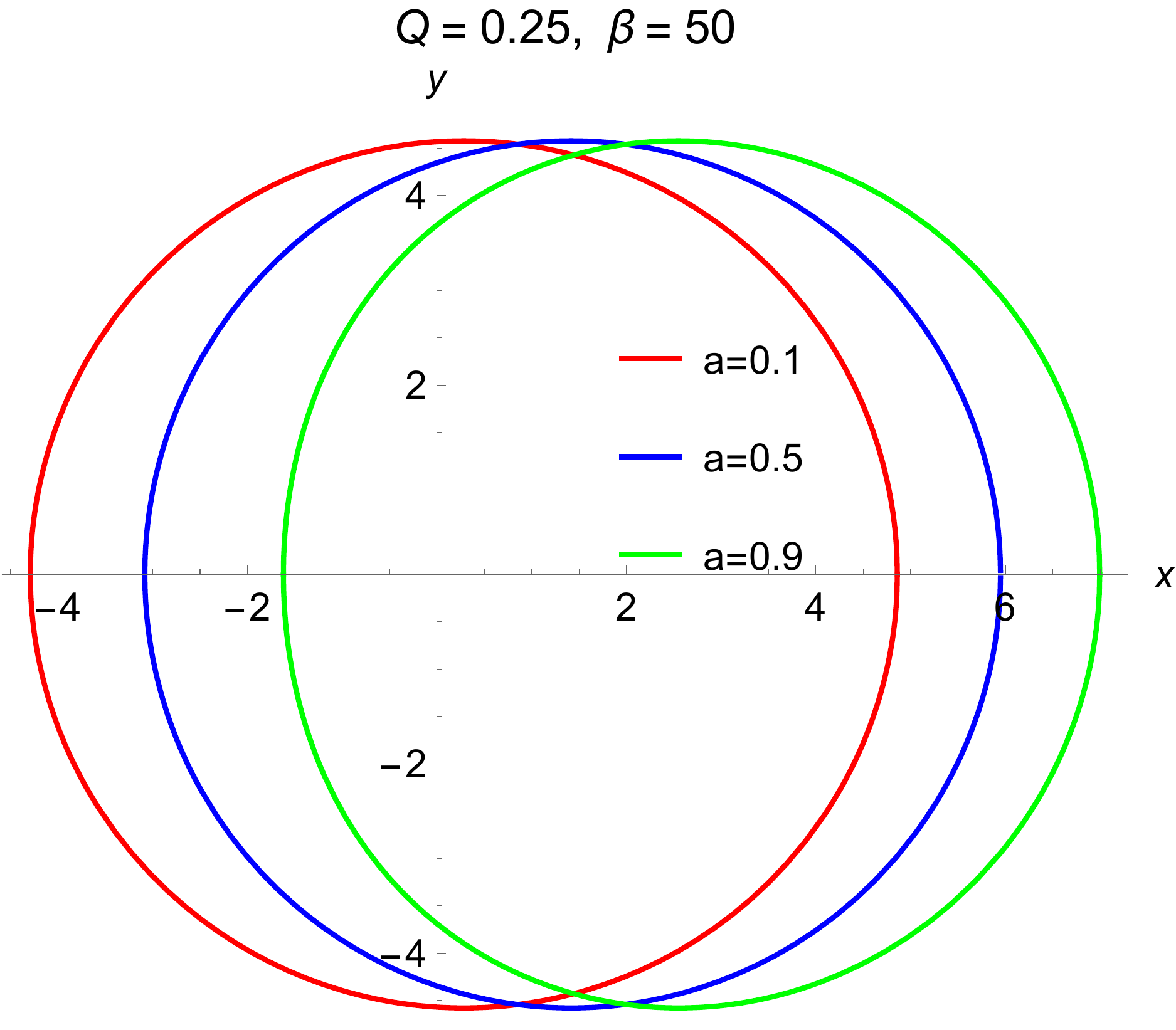}}
~
\subfigure{
\includegraphics[height=5.09cm,width=5.6cm]{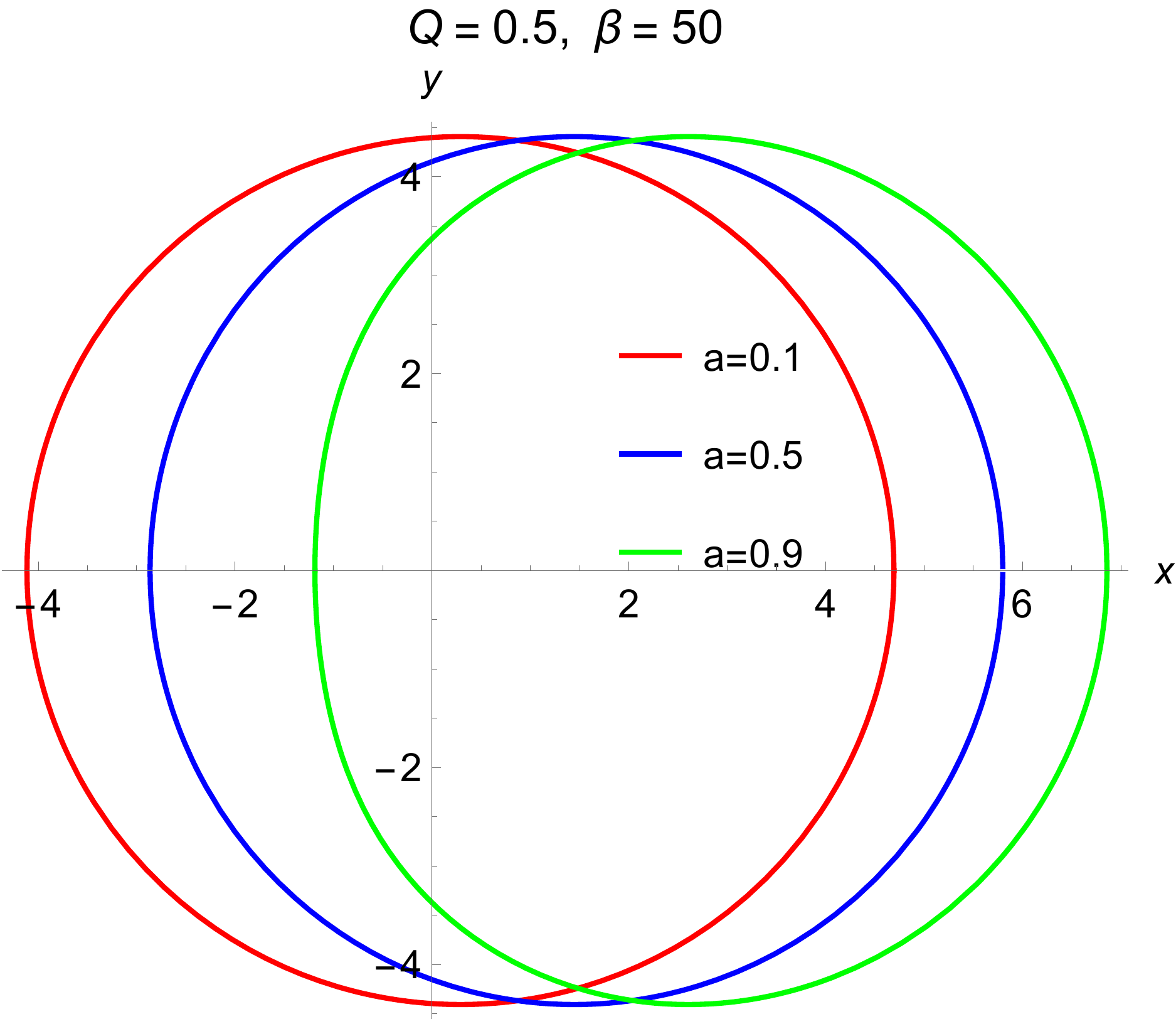}}
\subfigure{
\includegraphics[height=5.79cm,width=5.6cm]{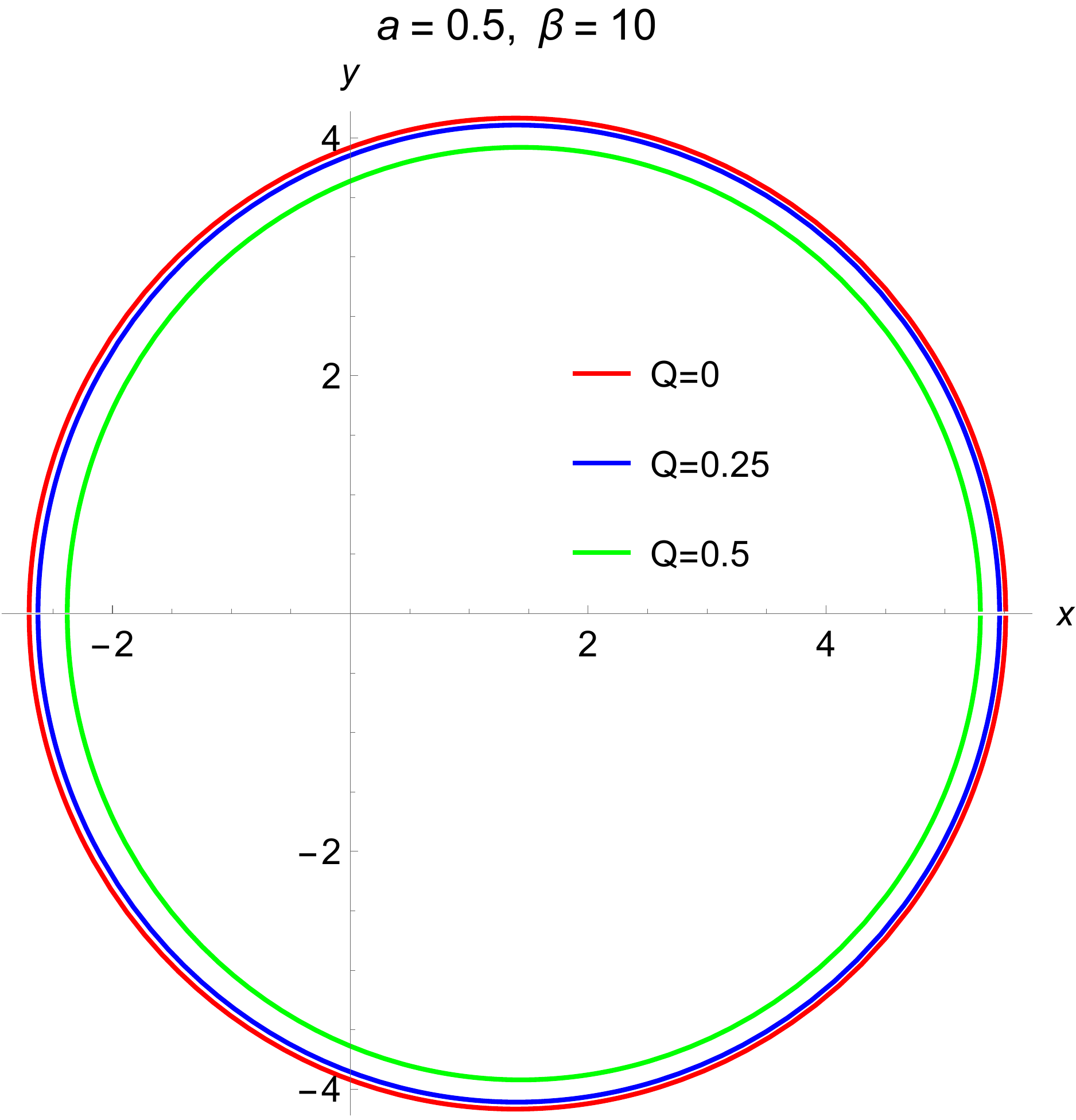}}
~
\subfigure{
\includegraphics[height=5.96cm,width=5.6cm]{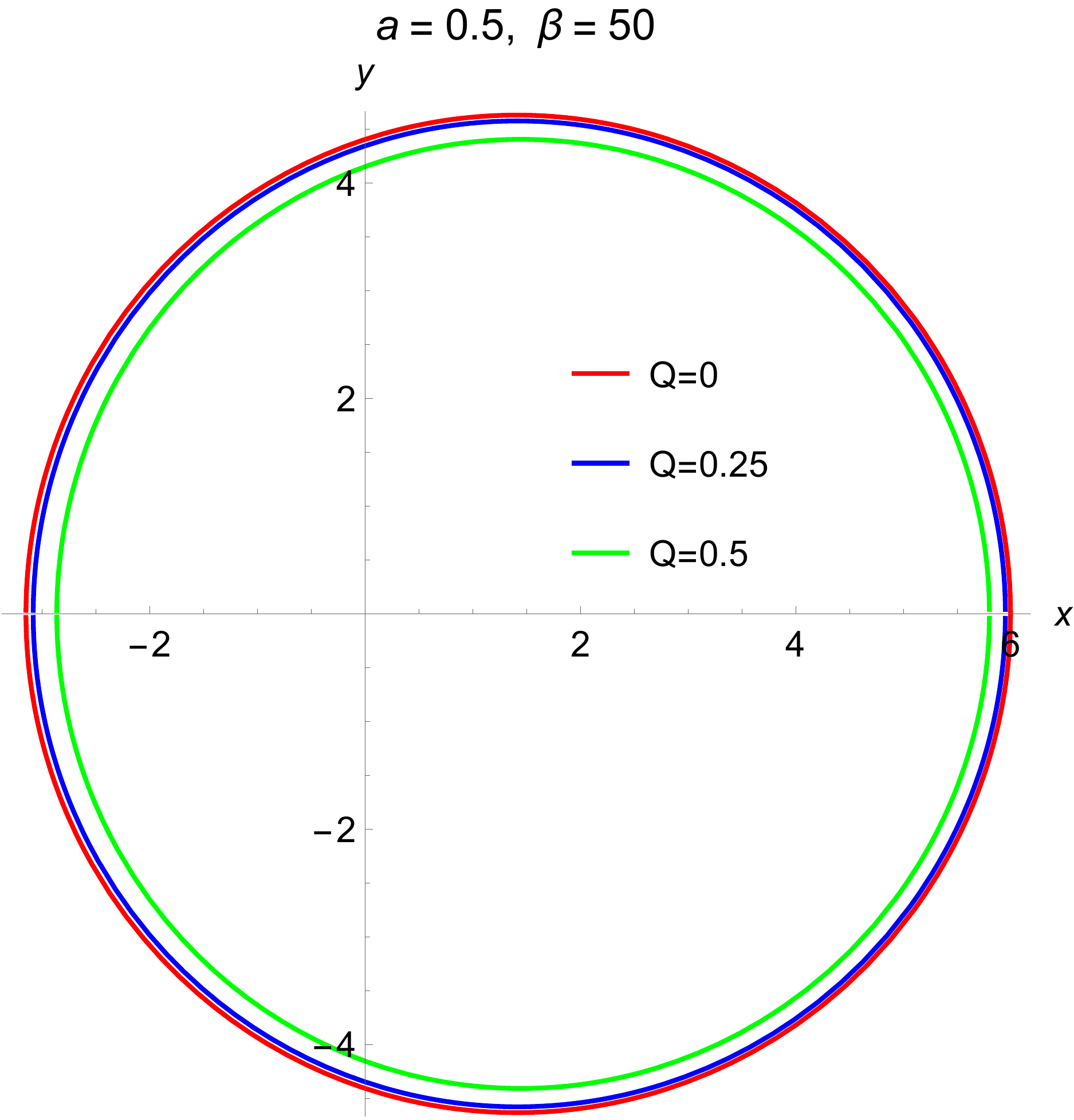}}
~
\subfigure{
\includegraphics[height=5.87cm,width=5.6cm]{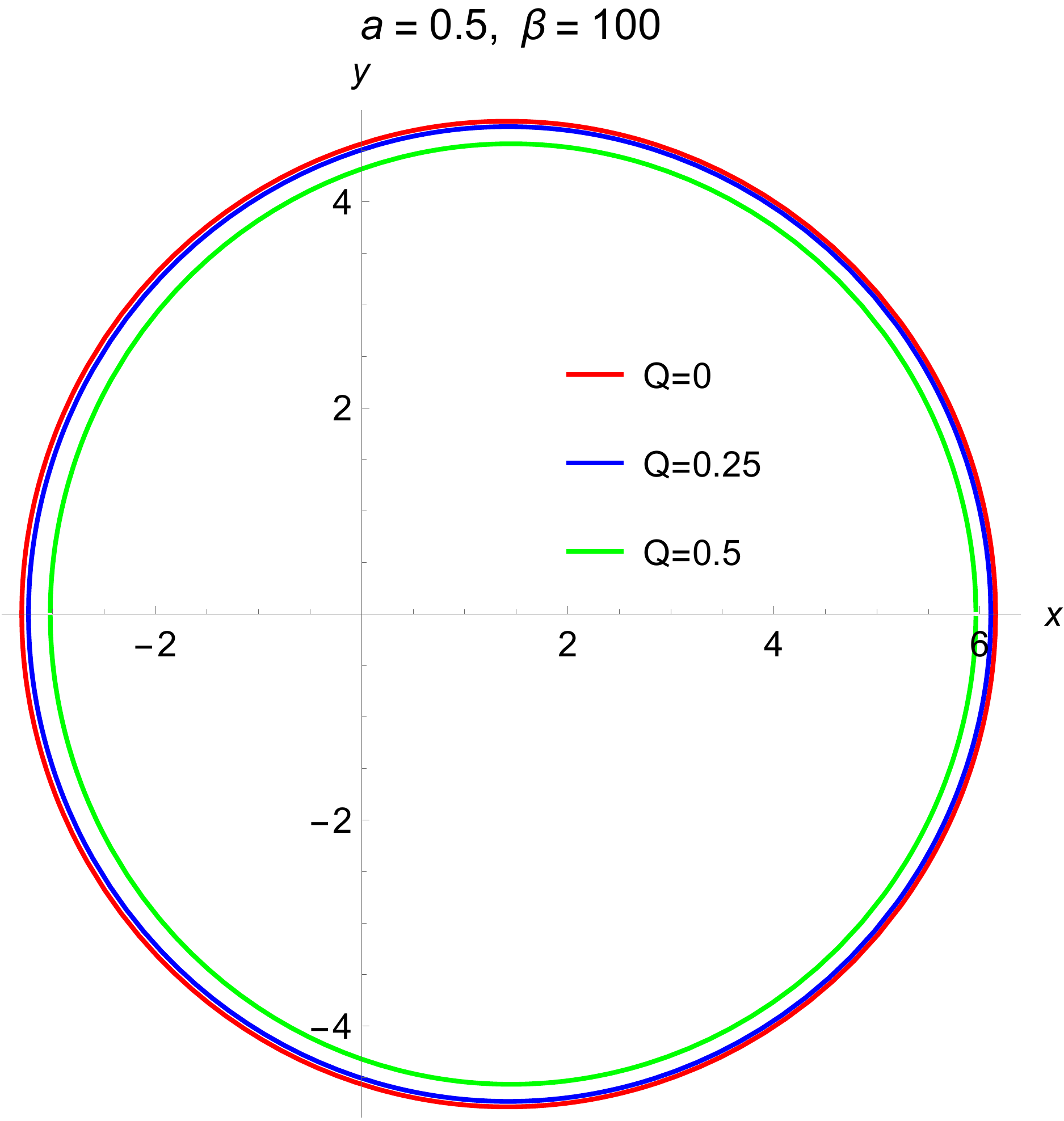}}
\subfigure{
\includegraphics[height=5.88cm,width=5.6cm]{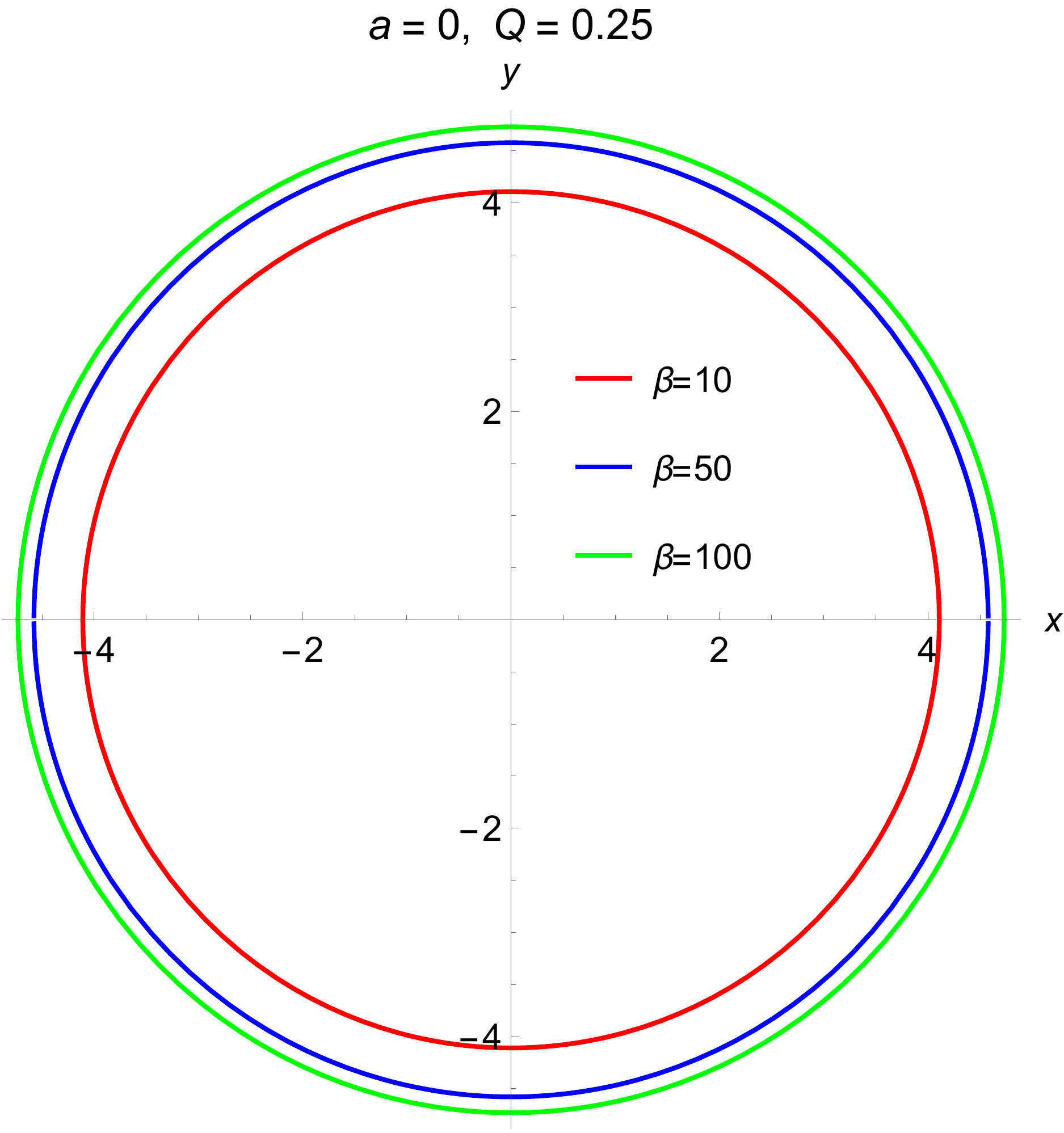}}
~
\subfigure{
\includegraphics[height=5.84cm,width=5.6cm]{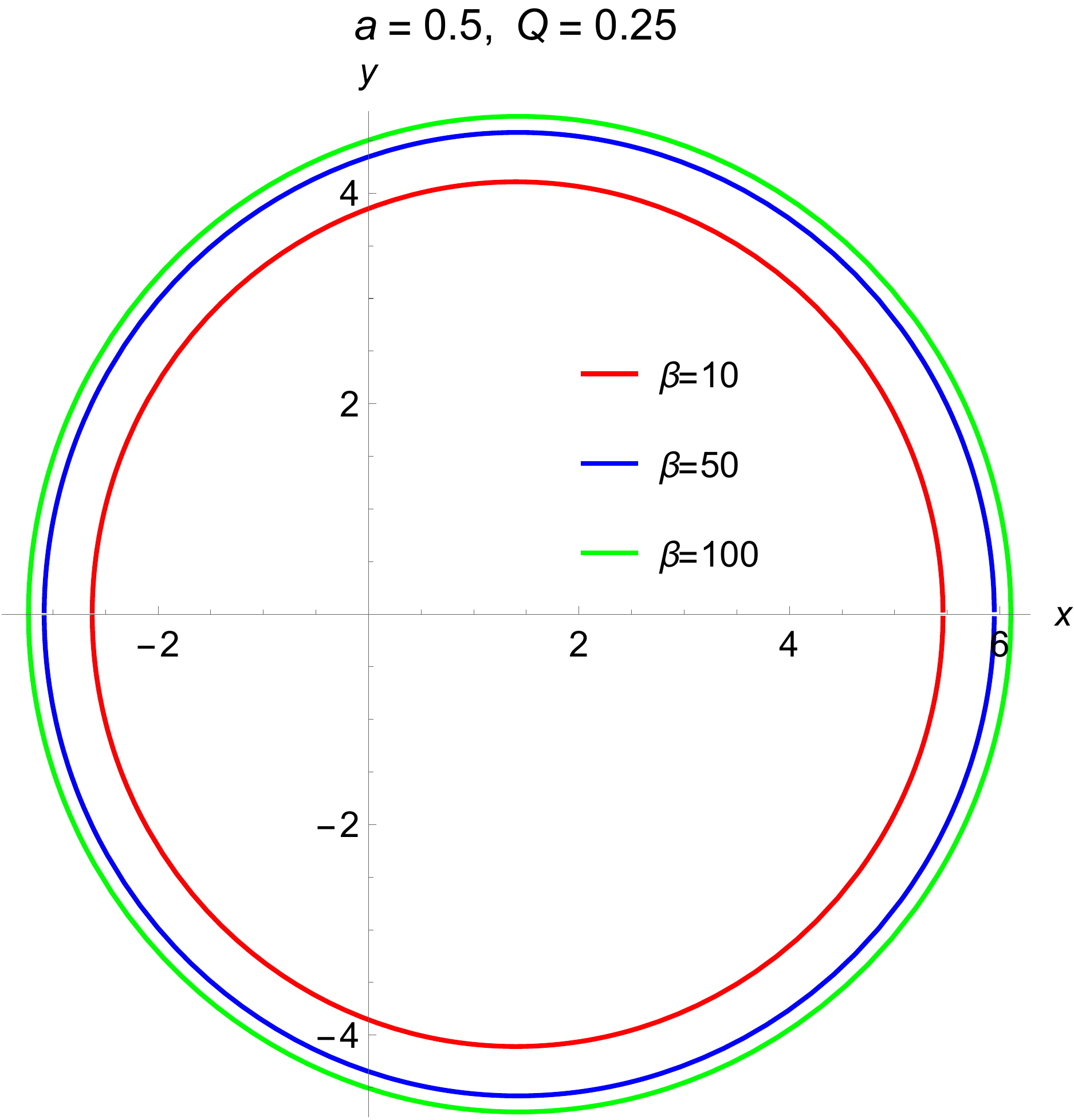}}
~
\subfigure{
\includegraphics[height=5.96cm,width=5.6cm]{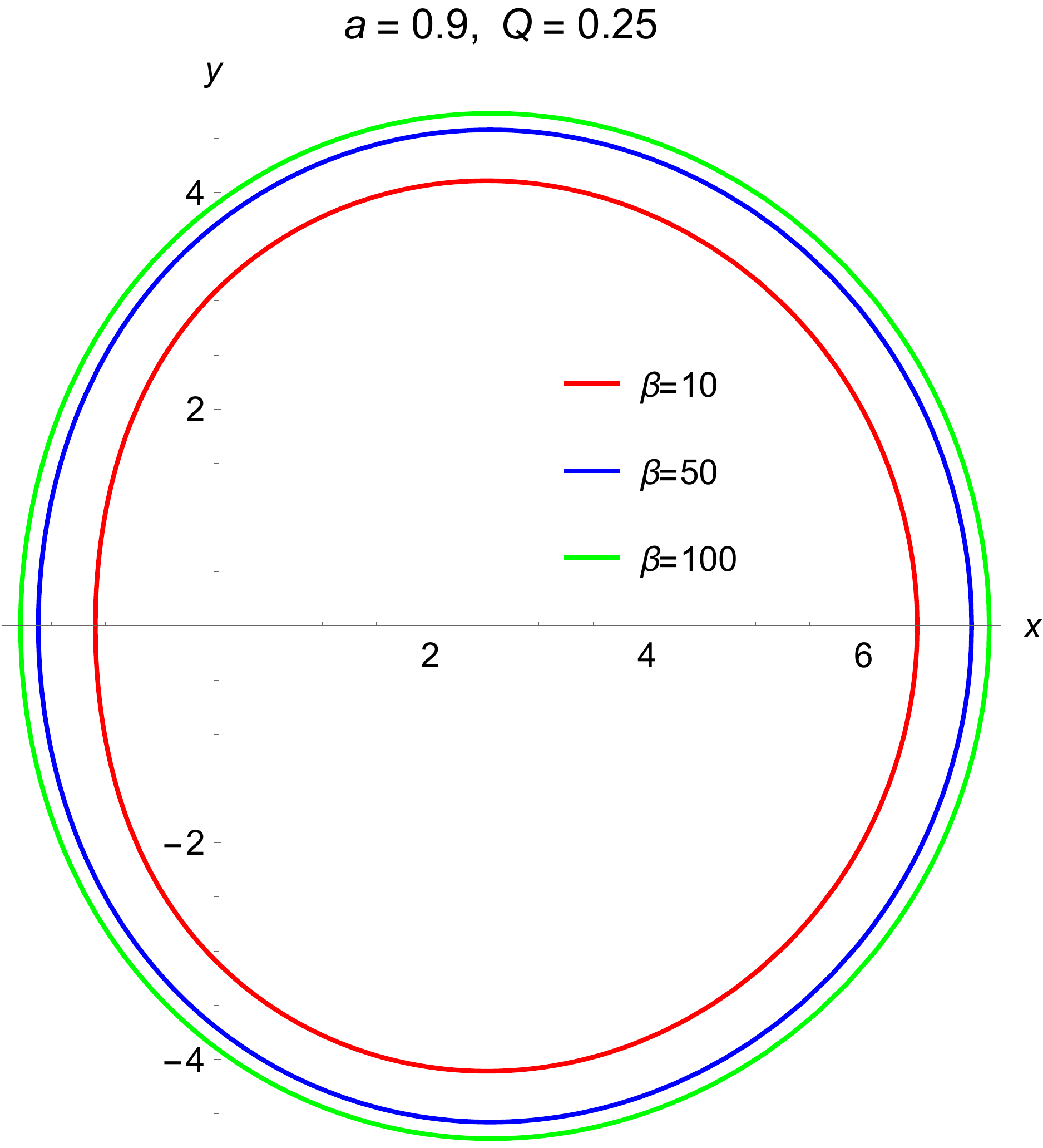}}
\end{center}
\caption{Stereographic projection of shadows in strong field w.r.t a remote equatorial observer. The top panel corresponds to the curves for $a=0.1$, $0.5$ and $0.9$. The middle panel corresponds to the curves for $Q=0$, $0.25$ and $0.5$. The bottom panel corresponds to the curves for $\beta=10$, $50$ and $100$.}
\end{figure}
\\

\subsection{Distortion in Strong Field}
So far, we have studied the shapes of shadow images. In order to quantify the actual size of the shadow, we need to determine the value of an observable that is equivalent to the linear radius of the shadow. It is denoted by $R_{sh}$ and is defined by \cite{32,75}
\begin{equation}
R_{sh}=\frac{(x_t-x_r)^2+y_t^2}{2|x_t-x_r|}. \label{73}
\end{equation}
It is the radius of an imaginary circle that touches the shadow image at three different points at the top with coordinates $(x_t,y_t)$, at the bottom with coordinates $(x_b,y_b)$ and the right most point with coordinates $(x_r,0)$. Here, $x$ and $y$ are the coordinates of the points on the stereographic shadow image and the subscripts $t$, $r$ and $b$ correspond to the top, the right most and the bottom point, respectively. We can locate these points on the shadow image as given in Fig. $\textbf{9}$ in \cite{75}. The above formula is useful for rotating BHs since for the non-rotating BHs, the imaginary circle will also touch the fourth point $(-x_r,0)$. Thus, the distance of curves from origin will be equal at all points. Hence, the radius of the shadow is same as the radius of the circle itself. The above value of $R_{sh}$ is useful in calculating the distortion and energy emission rates. The distortion actually characterizes the difference of the shadow images as a result of the flatness due to spin on one side. It can be calculated by the following relation:
\begin{equation}
\delta=\frac{|\bar{x}_l-x_l|}{R_{sh}}, \label{74}
\end{equation}
where $(x_l,0)$ is the point where shadow curve intersects the $-x$-axis and $(\bar{x}_l,0)$ is the point where imaginary circle intersects the $-x$-axis. Here, the subscript $l$ corresponds to the left most point on the shadow image, whereas, $\bar{x}$ corresponds to the point on the imaginary circle which is not coinciding the shadow curve. The distortion corresponding to shadow curves in strong field is denoted by $\delta_s$. The distortion is plotted in Fig. $\textbf{6}$. The upper panel depicts the distortion plots in strong field limit for an observer near the BH and the lower panel shows the distortion in strong field when the observer is far away from the BH. The first image in upper panel shows that the distortion increases with increasing spin value. The distortion decreases with increase in $\beta$ and this decreasing rate increases with increase in spin. There is no distortion for $a=0$. The second plot shows that distortion decreases with increase in $\beta$ while it increases with increasing charge $Q$. The third plot says that the distortion is almost negligible for least value of spin and increases with increasing spin $a$. Also, the distortion increases by increasing the value of charge. The lower panel depicts almost the same behaviour as the upper panel. However, the plots are only different in terms of the plot ranges and numbers. This is because of the fact that the shadow is observed at the equator at two different locations. The overall behavior of the plots show that the distortion decreases with increase in $\beta$. It means that the nonlinear electrodynamics is responsible for diminishing the flatness of the shadows. The increasing charge also increases the distortion. The spin has very prominent effect on the shadows and hence can be seen from the distortion pictures. The spin flatness is diminished by $\beta$ but still the distortion increases with increase in spin.
\begin{figure}[t]
\begin{center}
\subfigure{
\includegraphics[height=3.73cm,width=5.6cm]{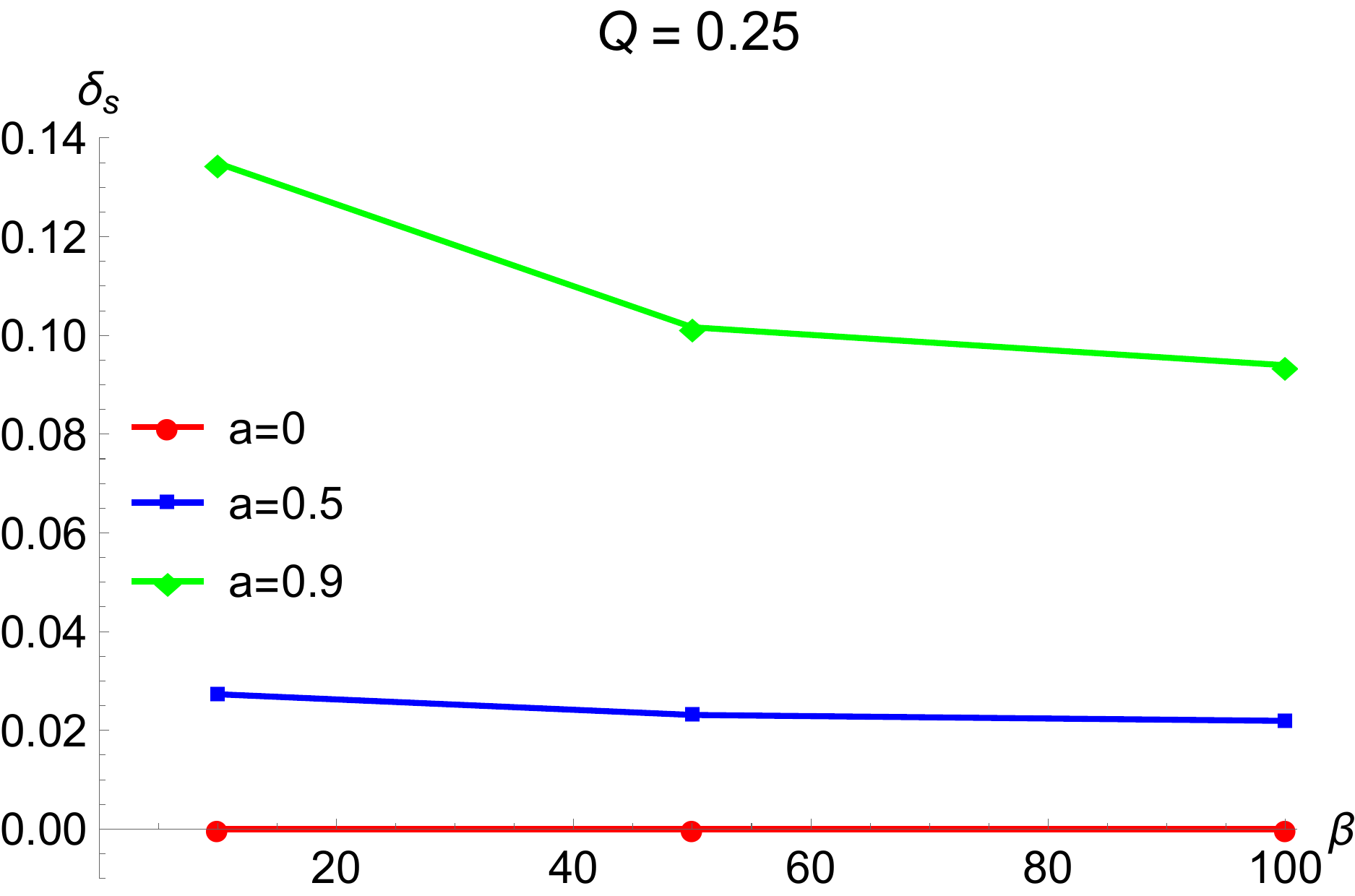}}
~
\subfigure{
\includegraphics[height=3.73cm,width=5.6cm]{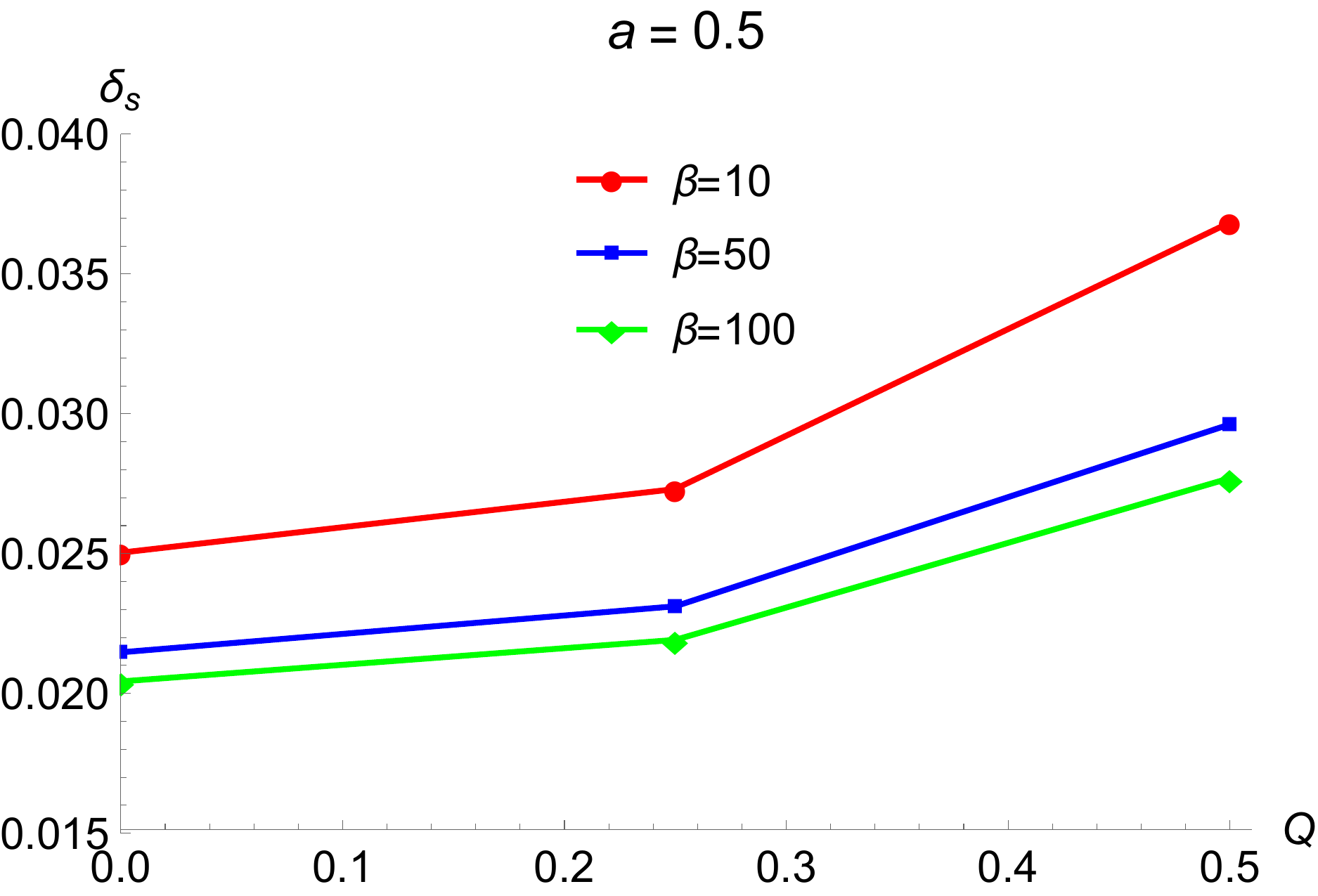}}
~
\subfigure{
\includegraphics[height=3.73cm,width=5.6cm]{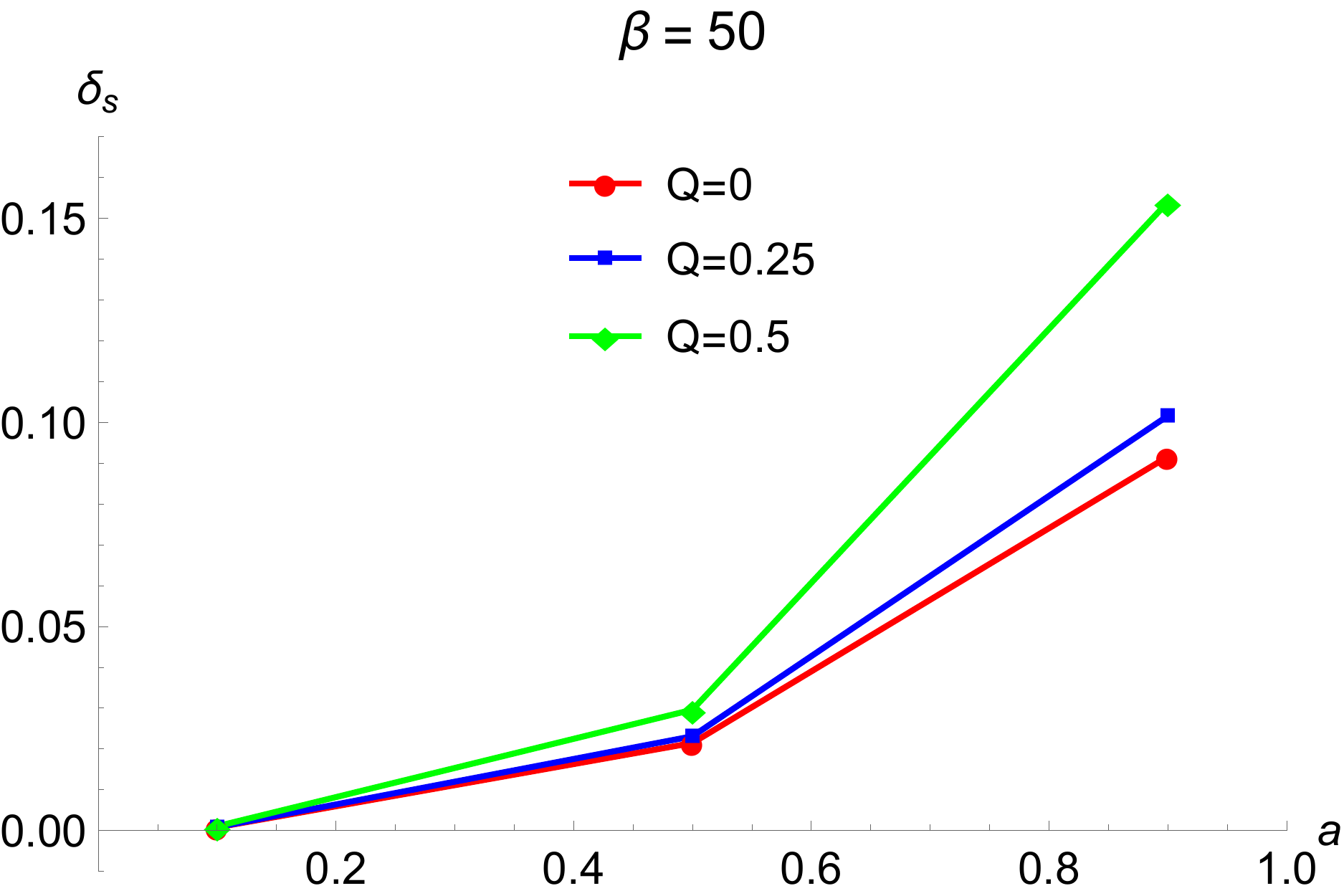}}
\subfigure{
\includegraphics[height=3.73cm,width=5.6cm]{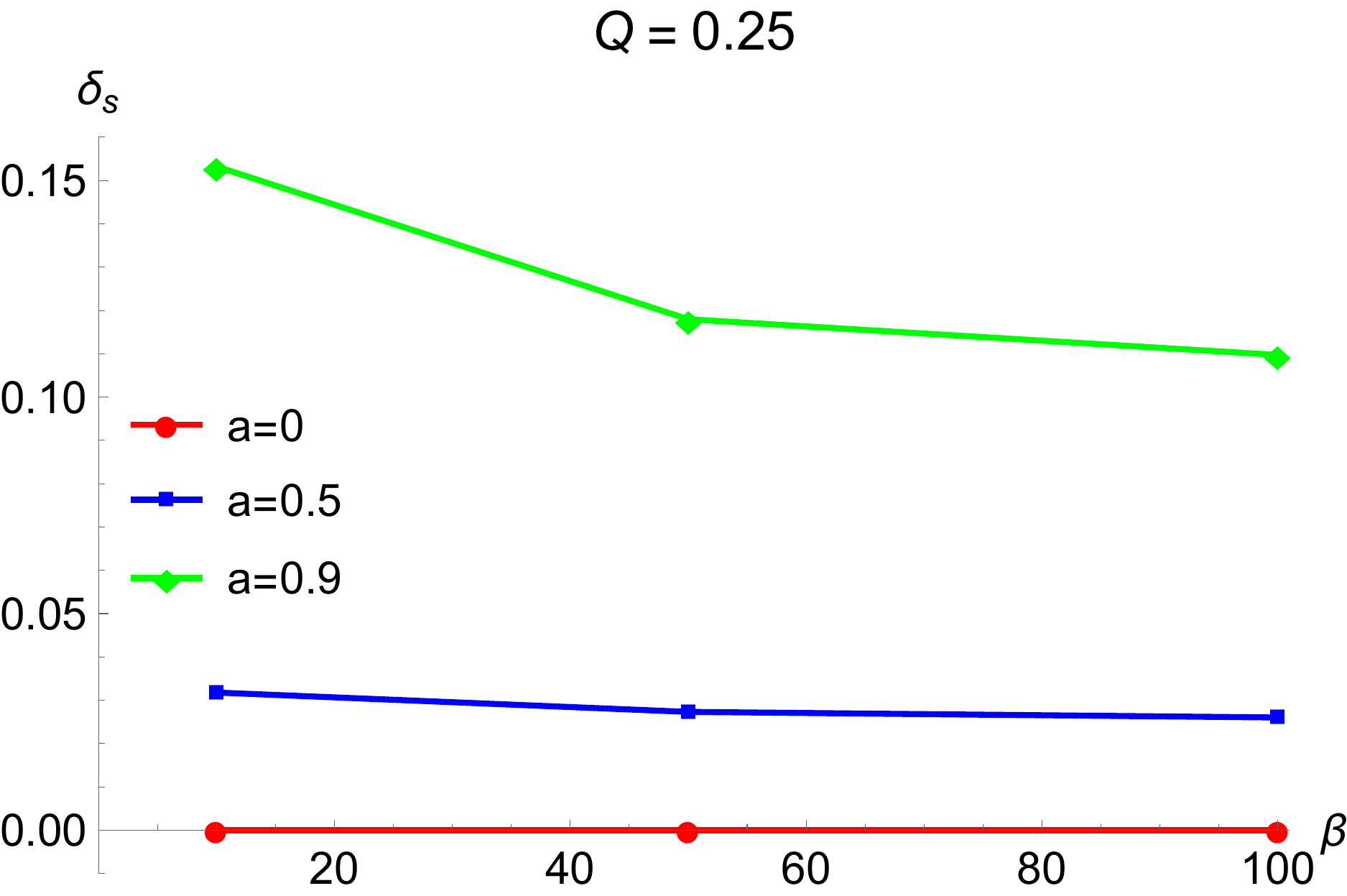}}
~
\subfigure{
\includegraphics[height=3.73cm,width=5.6cm]{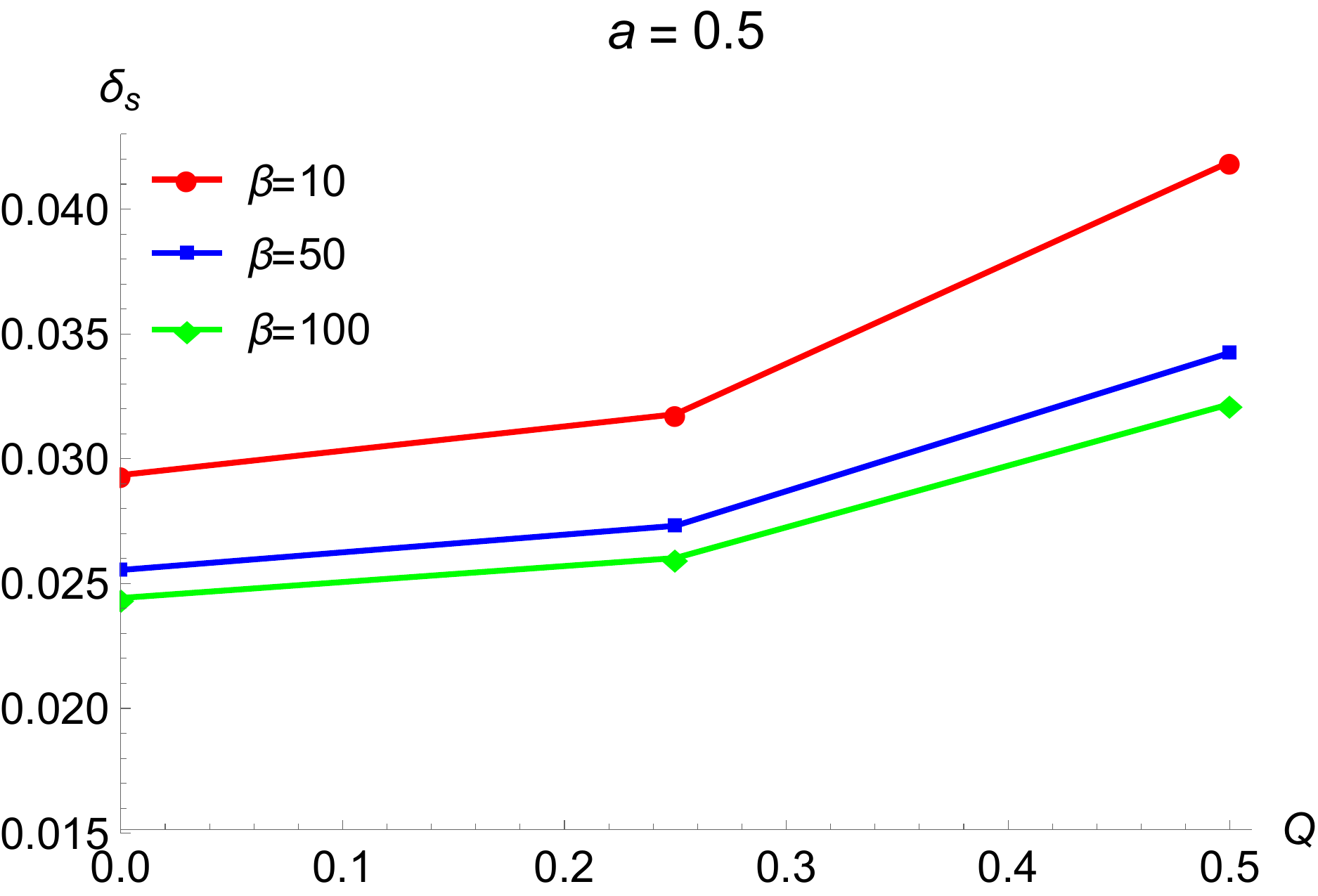}}
~
\subfigure{
\includegraphics[height=3.73cm,width=5.6cm]{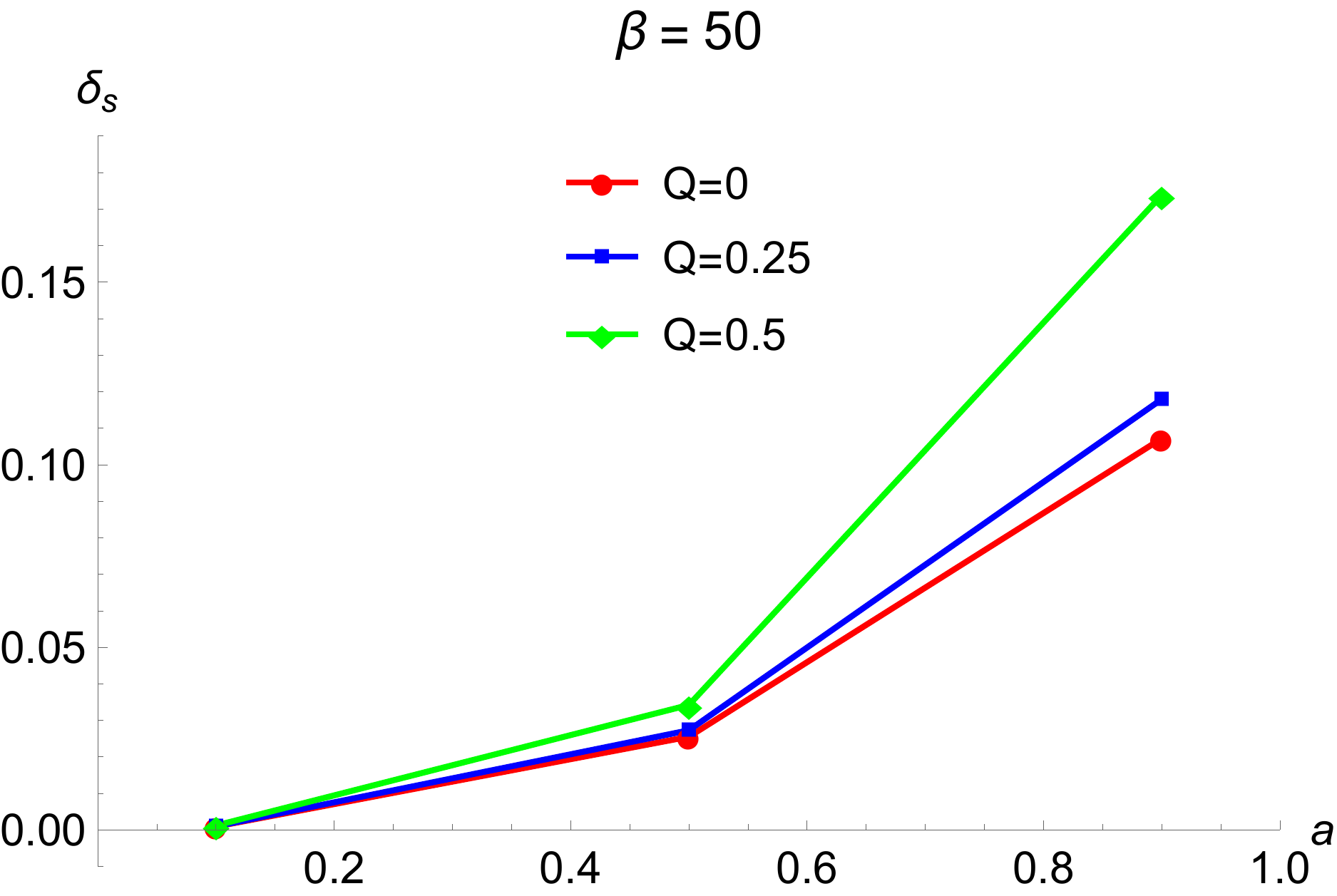}}
\end{center}
\caption{Plots for distortion in strong field for a nearer observer (upper panel) and remote observer (lower panel) w.r.t $\beta$, $Q$ and $a$.}
\end{figure}

\subsection{Energy Emission Rate}
As we know that particles are created and annihilated near the horizon due to the quantum fluctuations inside the BH. The particles possessing positive energy will fly off from the BH by the effect called quantum tunneling, which causes the BH to evaporate and the associated energy will also be emitted \cite{49}. The probability of any absorption process is measured by absorption cross section. From the astrophysical point of view there is no interest in the measurements by an observer located near the horizon. Hence, at large distances, the high energy absorption cross section has a correspondence with the BH shadow. At such high scale of energy, the value of BH's absorption cross section remains around a limit denoted as $\sigma_{lim}$ which is a constant value. This value of $\sigma_{lim}$ was found equal to the geometrical area of the photon sphere \cite{75,76,77,78}. Since the interface of photon sphere and the shadow is common, hence it can be related to the absorption cross section as
\begin{equation}
\sigma_{lim}\approx\pi R_{sh}^2. \label{75}
\end{equation}
The relation for energy emission rate is given as
\begin{equation}
\mathcal{E}_{\omega t}=\frac{d^2\mathcal{E}(\omega)}{d\omega dt}=\frac{2\pi^2\sigma_{lim}\omega^3}{e^{\omega/T_H}-1}\approx\frac{2\pi^3R_{sh}^2\omega^3}{e^{\omega/T_H}-1}, \label{76}
\end{equation}
where $\omega$ is the frequency, $T_H=\bar{\kappa}/2\pi$ is the Hawking temperature and
\begin{equation}
\bar{\kappa}=\frac{\Delta'_s}{2(r^2+a^2)}\bigg|_{r=r_+} \label{77}
\end{equation}
is the surface gravity at the outer horizon $r_+$. For $a=0$,
\begin{equation}
\bar{\kappa}=\frac{1}{2}f'(r)_s\bigg|_{r=r_+} \label{78}
\end{equation}
is the surface gravity. The behavior of emission rate for different cases is plotted in Fig. $\textbf{7}$ which shows that the emission rate increases with increase in $a$, $\beta$ and $Q$ in all plots. It means that by increasing the charge, spin or $\beta$, the evaporation process also gets faster. Also it can be seen that the peak of emission rate is shifted towards left with increase in all parameters. However, for $a=0$, the emission profile is different from the rotating cases.

\section{Conclusion}
In this paper, we have considered a nonlinear electrodynamics BH and identified the effective metric in strong field limit that is converted into its rotating counterpart by applying the modified Newman-Janis algorithm. In the strong field approximation, we have studied the radius of photon sphere and horizon structure for non-rotating and rotating cases, respectively. Furthermore, the analytical scheme is developed for shadow study starting with Hamilton-Jacobi formalism and then by implying the method of orthonormal tetrads. The equations for celestial coordinates in the observer's sky are derived in strong field which, further, are inserted into stereographic projection coordinates in order to convert a spherical image into a $2$D image. Lastly, distortion and energy emission rate are discussed. The results presented with the description of figures are summarized below:\\
\begin{figure}[t]
\begin{center}
\subfigure{
\includegraphics[height=3.8cm,width=5.6cm]{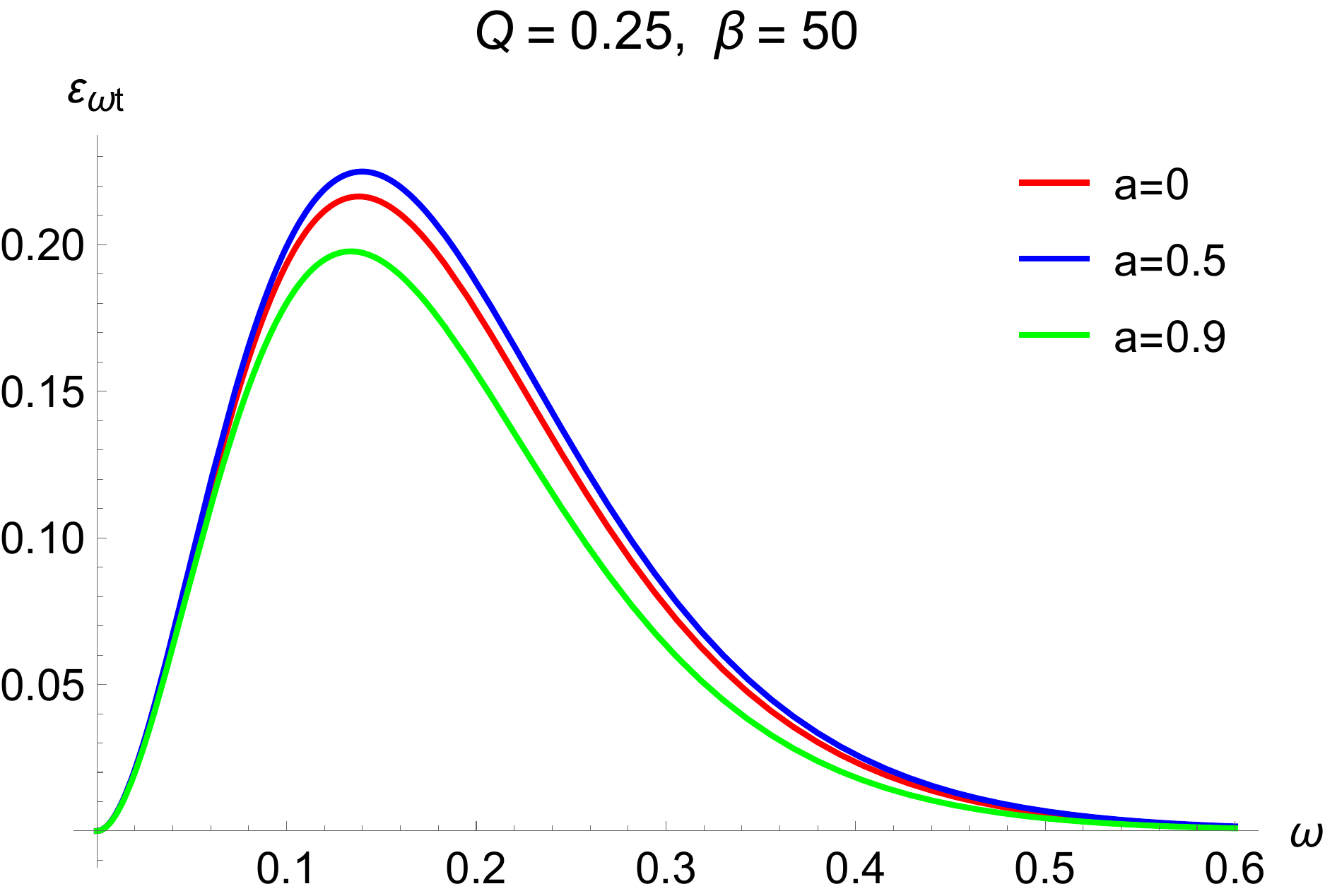}}
~
\subfigure{
\includegraphics[height=3.8cm,width=5.6cm]{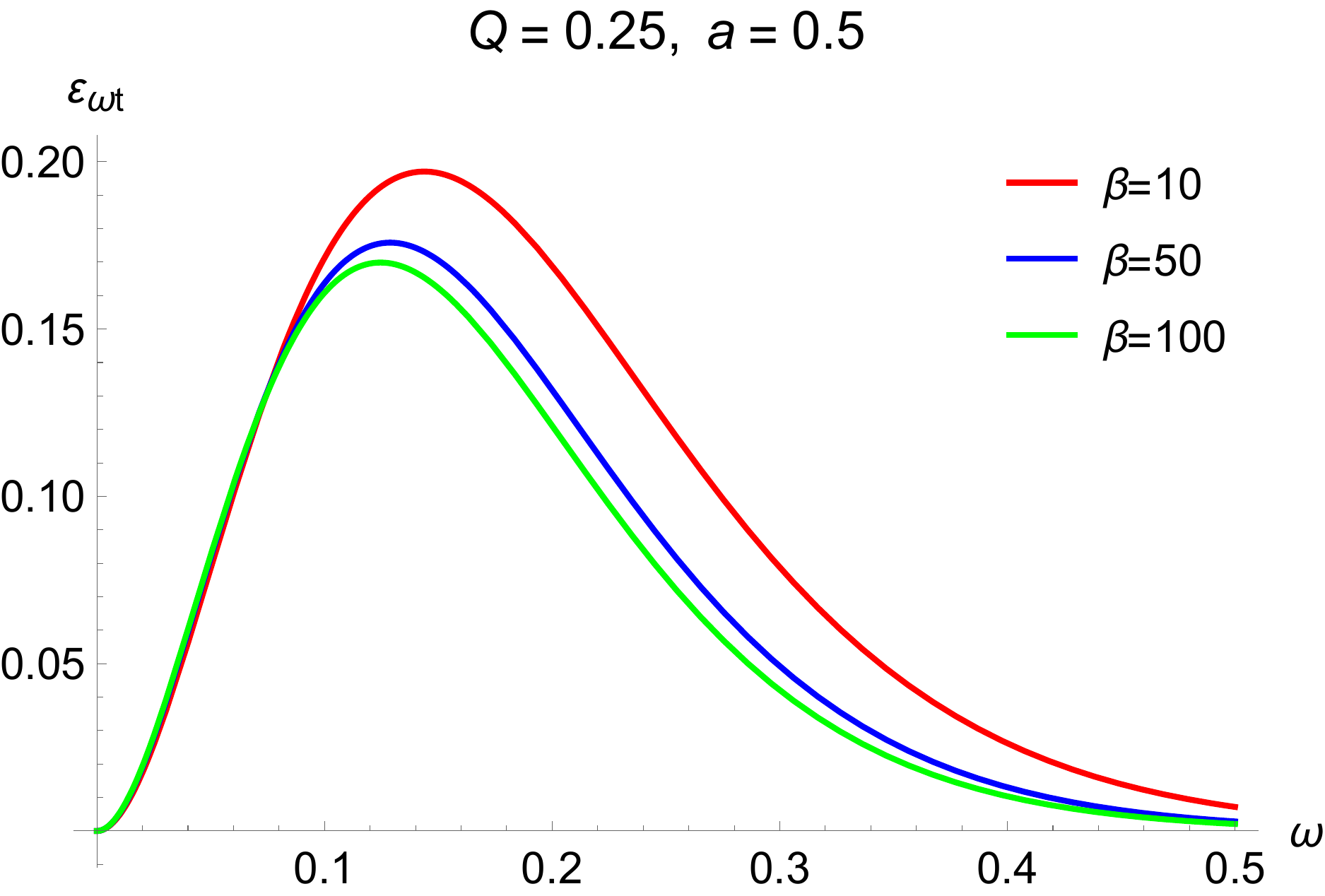}}
~
\subfigure{
\includegraphics[height=3.8cm,width=5.6cm]{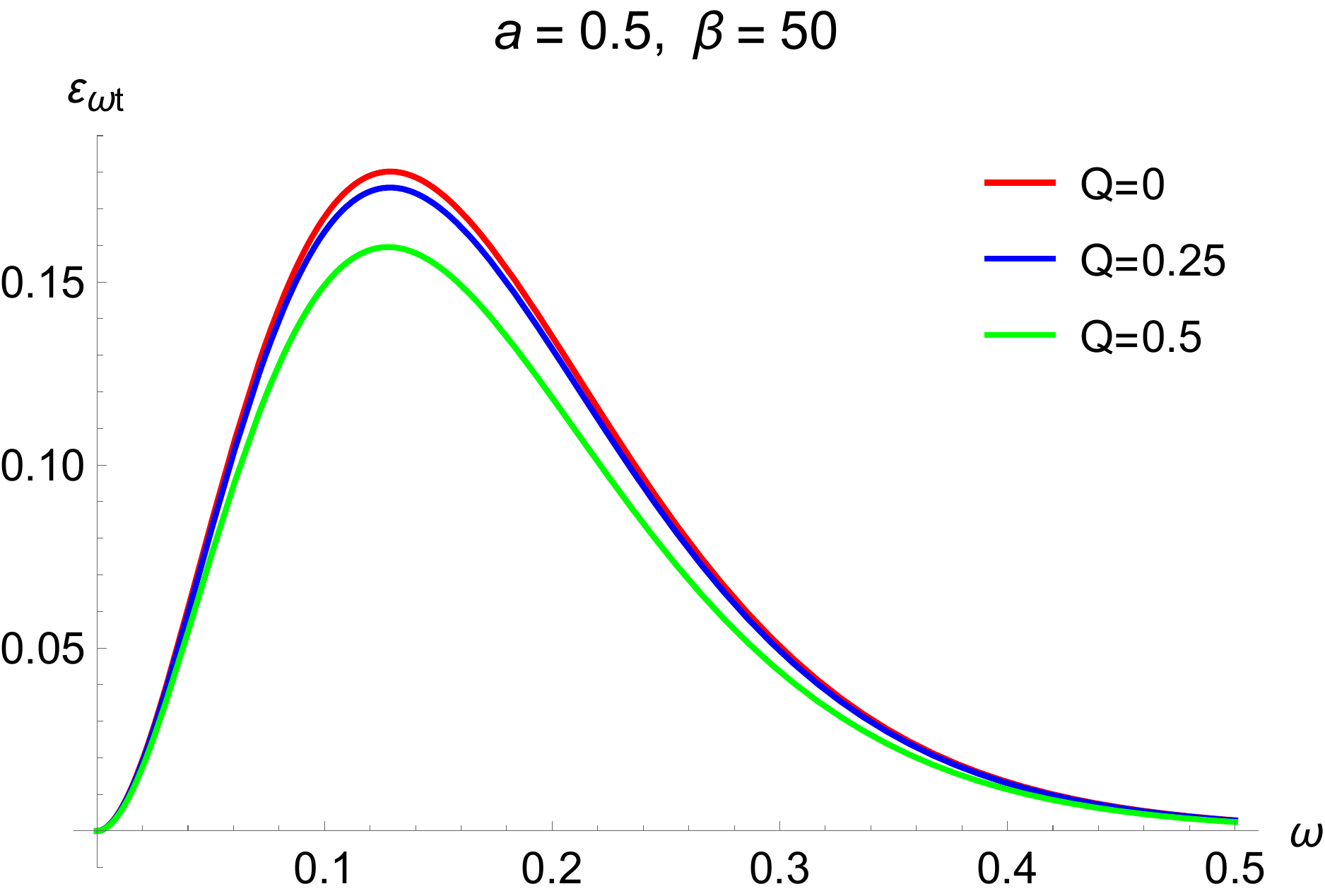}}
\end{center}
\caption{Plots showing the energy emission rate in strong field w.r.t frequency $\omega$ detected by a remote observer.}
\end{figure}

$\bullet$ The metric function in the strong field suggests that the nonlinear electrodynamics effects can not be removed in the vicinity of the BH. From Fig. $\textbf{1}$, it can be seen that the size of photon sphere increases with increasing $\beta$ and decreases with increase in $Q$. It also shows that the results for Reissner-Nordstr\"{o}m metric and the metric (\ref{8}) are different from each other. For particular values of $C$ and $\beta$, the metric (\ref{8}) reduces into Reissner-Nordstr\"{o}m metric. \\

$\bullet$ Figure $\textbf{2}$ corresponds to the horizon radius. We can see that the horizon radius increases with increasing $\beta$ and central singularity is formed only for the last case in the left panel. The other two pictures show that there is only one horizon for certain values of spin. From the right panel, we can say that the horizon becomes short by increasing the charge and the Cauchy horizon is formed for every value of spin only for one particular value of charge. The dashed curves show the horizon radii for Reissner-Nordstr\"{o}m BH and its rotating counterpart. It is obvious that these results are different from the results achieved in our study. \\

$\bullet$ Shadows for a nearer observer are plotted in Fig. $\textbf{3}$. It is found that the spin shifts the shadow to the right and a possible flatness is observed which is removed by the effect of nonlinear electrodynamics. The increasing charge decreases the size of shadow and the increasing $\beta$ increases the size of shadow. Fig. $\textbf{4}$ shows that as the observer moves away, the shadow size is reduced as expected. From Fig. $\textbf{5}$, we observe the same behaviour of the shadows as compared to the shadows in Fig. $\textbf{3}$ when the observer moves far away from the BH. The only measured difference is the size of the shadow as expected. \\

$\bullet$ The difference in shadows due to effect of spin and $\beta$ is observed in Fig. $\textbf{6}$. For higher values of $\beta$, the distortion decreases and it increases with increase in charge and spin. Lastly, the Fig. $\textbf{7}$ shows that the evaporation of the BH becomes slower by increasing charge, spin and $\beta$. The evaporation rate for non-rotating case is different from the rotating cases. \\

We conclude that the presence of nonlinear electrodynamics may strengthen the gravity due to the fact that the size of photon sphere, horizons and shadows increase with involvement of $\beta$. Also, the central singularity is not formed for some cases in the presence of $\beta$ and other parameters. Moreover, the capture of photon, horizon and the shadow variation with the charge suggests the possibility of reduction of the strength of gravity due to involvement of charge. The effect of spin on the shadows can be removed by introducing nonlinearity of electromagnetic field. It can also be concluded that the Reissner-Nordstr\"{o}m BH and its rotating counterpart are the special limiting cases of the strong field metric.


\begin{thebibliography}{36}
\bibitem{1} F. W. Dyson, A. S. Eddington and C. Davidson, Phil. Trans. Royal Soc. A $\mathbf{220}$, $0009$ $(1920)$.
\bibitem{2} P. Schneider, J. Ehlers and E. E. Falco, \textit{Gravitational Lenses}, (Springer-Verlag, Berlin, $1992$), ISBN $978-3-662-03758-4$.
\bibitem{3} R. D. Blandford and R. Narayan, Ann. Rev. Astron. Astrophys. $\mathbf{30}$, $311$ $(1992)$.
\bibitem{4} A. O. Petters, H. Levine and J. Wambsganss, \textit{Singularity Theory and Gravitational Lensing}, (Birkh$\ddot{a}$user, Boston, MA, $2001$), ISBN $978-1-4612-0145-8$.
\bibitem{5} V. Perlick, Liv. Rev. Relativity $\mathbf{7}$, $9$ $(2004)$.
\bibitem{6} M. Bartelmann, Class. Quantum Grav. $\mathbf{27}$, $233001$ $(2010)$.
\bibitem{7} J. Wambsganss, Liv. Rev. Relativity $\mathbf{1}$, $12$ $(1998)$.
\bibitem{8} F. Atamurotov, A. Abdujabbarov and W.-B. Han, Phys. Rev. D $\mathbf{104}$, $084015$ $(2021)$.
\bibitem{9} A. Abdujabbarov, B. Ahmedov, N. Dadhich and F. Atamurotov, Phys. Rev. D $\mathbf{96}$, $084017$ $(2017)$.
\bibitem{10} V. Perlick and O. Y. Tsupko, Phys. Reports $\mathbf{947}$, $1$ $(2022)$.
\bibitem{11} S. Chandrasekhar, \textit{The Mathematical Theory of Black Holes}, (Oxford University Press, $1983$), ISBN $0-19-851291-0$.
\bibitem{12} J.-P. Luminet, Astron. Astrophys. $\mathbf{75}$, $228$ $(1979)$.
\bibitem{13} S. U. Viergutz, Astron. Astrophys. $\mathbf{272}$, $355$ $(1993)$.
\bibitem{14} J. L. Synge, Mon. Not. Roy. Astron. Soc. $\mathbf{131}$, $463$ $(1966)$.
\bibitem{15} Y.B. Zeldovich and I.D. Novikov, Usp. Fiz. Nauk $\mathbf{86}$, $447$ $(1965)$.
\bibitem{16} P. J. Young, Phys. Rev. D $\mathbf{14}$, $3281$ $(1976)$.
\bibitem{17} I. G. Dymnikova, Sov. Phys. Usp. $\mathbf{29}$, $215$ $(1986)$.
\bibitem{18} A. Grenzebach, V. Perlick and C. Lammerzahl, Phys. Rev. D $\mathbf{89}$, $124004$ $(2014)$.
\bibitem{19} A. Grenzebach, V. Perlick and C. Lammerzahl, Int. J. Mod. Phys. D $\mathbf{24}$, $1542024$ $(2015)$.
\bibitem{20} J. M. Bardeen, Gordon and Breach, Science Publishers, New York, $23$rd Summer school of theoretical physics, Les Houches, France, $215$ $(1973)$.
\bibitem{21} M. Sharif and S. Iftikhar, Eur. Phys. J. C $\mathbf{76}$, $630$ $(2016)$.
\bibitem{22} H.-M. Wang, Y.-M. Xu and S.-W. Wei, J. Cos. Ast. Phys. $\mathbf{03}$, $046$ $(2019)$.
\bibitem{23} A. Das, A. Saha and S. Gangopadhyay, Eur. Phys. J. C $\mathbf{80}$, $180$ $(2020)$.
\bibitem{24} K. Jusufi, M. Jamil, P. Salucci, T. Zhu, and S. Haroon, Phys. Rev. D $\mathbf{100}$, $044012$ $(2019)$.
\bibitem{25} F. Atamurotov, U. Papnoi and K. Jusufi, Class Quant. Grav. $\mathbf{39}$, $025014$ $(2021)$.
\bibitem{26} S.-W. Wei and Y.-X. Liu, Eur. Phys. J. Plus $\mathbf{136}$, $436$ $(2021)$.
\bibitem{27} S.-W. Wei, Y.-C. Zou, Y.-X. Liu and R. B. Mann, J. Cos. Ast. Phys. $\mathbf{08}$, $030$ $(2019)$.
\bibitem{28} B.-H. Lee, W. Lee and Y. S. Myung, Phys. Rev. D $\mathbf{103}$, $064026$ $(2021)$.
\bibitem{29} R. Kumar and S. G. Ghosh, J. Cos. Ast. Phys. $\mathbf{07}$, $053$ $(2020)$.
\bibitem{30} X. Hou, Z. Xu, M. Zhou and J. Wang, J. Cos. Ast. Phys. $\mathbf{7}$, $015$ $(2018)$.
\bibitem{31} H. Falcke, F. Melia and E. Agol, Astrophys. J. $\mathbf{528}$, $L13$ $(2000)$.
\bibitem{32} K. Hioki and K. Maeda, Phys. Rev. D $\mathbf{80}$, $024042$ $(2009)$.
\bibitem{33} C. Bambi and K. Freese, Phys. Rev. D $\mathbf{79}$, $043002$ $(2009)$.
\bibitem{34} L. Amarilla, E. F. Eiroa and G. Giribet, Phys. Rev. D $\mathbf{81}$, $124045$ $(2010)$.
\bibitem{35} A. Abdujabbarov, F. Atamurotov, Y. Kucukakca, B. Ahmedov and U. Camci, Astrophys. Space. Sci. $\mathbf{344}$, $429$ $(2013)$.
\bibitem{36} L. Amarilla and E. F. Eiroa, Phys. Rev. D $\mathbf{85}$, $064019$ $(2012)$.
\bibitem{37} L. Amarilla and E. F. Eiroa, Phys. Rev. D $\mathbf{87}$, $044057$ $(2013)$.
\bibitem{38} A. Abdujabbarov, M. Amir, B. Ahmedov and S. G. Ghosh, Phys. Rev. D $\mathbf{93}$, $104004$ $(2016)$.
\bibitem{39} F. Atamurotov, A. Abdujabbarov and B. Ahmedov, Phys. Rev. D $\mathbf{88}$, $064004$ $(2013)$.
\bibitem{40} X. Hou, Z. Xu and J. Wang, J. Cos. Ast. Phys. $\mathbf{12}$, $040$ $(2018)$.
\bibitem{41} M. Amir, B. P. Singh and S. G. Ghosh, Eur. Phys. J. C $\mathbf{78}$, $399$ $(2018)$.
\bibitem{42} U. Papnoi, F. Atamurotov, S. G. Ghosh and B. Ahmedov, Phys. Rev. D $\mathbf{90}$, $024073$ $(2014)$.
\bibitem{43} F. Atamurotov, B. Ahmedov and A. Abdujabbarov, Phys. Rev. D $\mathbf{92}$, $084005$ $(2015)$.
\bibitem{44} N. Tsukamoto, Phys. Rev. D $\mathbf{97}$, $064021$ $(2018)$.
\bibitem{45} V. Perlick, O. Y. Tsupko and G. S. Bisnovatyi-Kogan, Phys. Rev. D $\mathbf{97}$, $104062$ $(2018)$.
\bibitem{46} P.-Z. He, Q.-Q. Fan, H.-R. Zhang and J.-B. Deng, Eur. Phys. J. C $\mathbf{80}$, $1195$ $(2020)$.
\bibitem{47} A. Övgün, İ. Sakallı and J. Saavedra, J. Cos. Ast. Phys. $\mathbf{10}$, $041$ $(2018)$.
\bibitem{48} K. Jusufi, M. Jamil and T. Zhu, Eur. Phys. J. C $\mathbf{80}$, $354$ $(2020)$.
\bibitem{49} U. Papnoi and F. Atamurotov, Phys. Dark Univ. $\mathbf{35}$, $100916$ $(2022)$.
\bibitem{50} S. Haroon, M. Jamil, K. Jusufi, K. Lin and R. B. Mann, Phys. Rev. D $\mathbf{99}$, $044015$ $(2019)$.
\bibitem{51} M. Born and L. Infeld, Proc. Roy. Soc. Lond. $\mathbf{144}$, $425$ $(1934)$.
\bibitem{52} M. Demianski, Found. Phys. $\mathbf{16}$, $187$ $(1986)$.
\bibitem{53} S. Fernando and D. Krug, Gen. Rel. Grav. $\mathbf{35}$, $129$ $(2003)$.
\bibitem{54} R.-G. Cai, D.-W. Pang and A. Wang, Phys. Rev. D $\mathbf{70}$, $124034$ $(2004)$.
\bibitem{55} O. Mi\v{s}kovi\'{c} and R. Olea, Phys. Rev. D $\mathbf{77}$, $124048$ $(2008)$.
\bibitem{56} S. I. Kruglov, Universe $\mathbf{4}$, $66$ $(2018)$.
\bibitem{57} S. I. Kruglov, Gen. Rel. Grav. $\mathbf{51}$, $9$ $(2019)$.
\bibitem{58} S. I. Kruglov, Eur. Phys. J. C $\mathbf{82}$, $292$ $(2022)$.
\bibitem{59} I. Gullu and S. H. Mazharimousavi, Phys. Scr. $\mathbf{96}$, $095213$ $(2021)$.
\bibitem{60} W. Javed, J. Abbas and Ali Övgün, Eur. Phys. J. C $\mathbf{79}$, $694$ $(2019)$.
\bibitem{61} S. I. Kruglov, Int. J. Geom. Meth. Mod. Phys. $\mathbf{12}$, $1550073$ $(2015)$.
\bibitem{62} A. Uniyal, R. C. Pantig and A. Övgün, arXiv:2205.11072 [gr-qc].
\bibitem{63} M. Azreg-Aïnou, Phys. Rev. D $\mathbf{90}$, $064041$ $(2014)$.
\bibitem{64} M. Azreg-Aïnou, Eur. Phys. J. C $\mathbf{74}$, $2865$ $(2014)$.
\bibitem{65} S. I. Kruglov, Ann. Phys. $\mathbf{353}$, $299$ $(2015)$.
\bibitem{66} E. T. Newman and A. I. Janis, J. Math. Phys. $\mathbf{6}$, $915$ $(1965)$.
\bibitem{67} E. T. Newman, E. Couch, K. Chinnapared, A. Exton, A. Prakash and R. Torrence, J. Math. Phys. $\mathbf{6}$, $918$ $(1965)$.
\bibitem{68} T. Johannsen and D. Psaltis, Phys. Rev. D $\mathbf{83}$, $124015$ $(2011)$.
\bibitem{69} K. Jusufi, M. Jamil, H. Chakrabarty, Q. Wu, C. Bambi and A. Wang, Phys. Rev. D $\mathbf{101}$, $044035$ $(2020)$.
\bibitem{70} C. Bambi and L. Modesto, Phys. Lett. B $\mathbf{721}$, $329$ $(2013)$.
\bibitem{71} D. Hansen and N. Yunes, Phys. Rev. D $\mathbf{88}$, $104020$ $(2013)$.
\bibitem{72} M. Fathi, M. Olivares and J. R. Villanueva, Galaxies $\mathbf{9}$, $43$ $(2021)$.
\bibitem{73} R. K. Walia, S. D. Maharaj and S. G. Ghosh, Eur. Phys. J. C $\mathbf{82}$, $547$ $(2022)$.
\bibitem{74} B. Carter, Phys. Rev. $\mathbf{174}$, $1559$ $(1968)$.
\bibitem{75} M. Amir and S. G. Ghosh, Phys. Rev. D $\mathbf{94}$, $024054$ $(2016)$.
\bibitem{76} S. Wei and Y. Liu, J. Cos. Ast. Phys. $\mathbf{11}$, $063$ $(2013)$.
\bibitem{77} B. Mashhoon, Phys. Rev. D $\mathbf{7}$, $2807$ $(1973)$.
\bibitem{78} Y. Décanini, G. Esposito-Farèse and A. Folacci, Phys. Rev. D $\mathbf{83}$, $044032$ $(2011)$.

\end{thebibliography}
\end{document}